\documentclass[sigconf, nonacm]{acmart}
\usepackage{amsmath}

\usepackage{balance}
\usepackage{graphicx}
\usepackage{framed}
\usepackage{color}
\usepackage{listings}
\usepackage{subcaption}
 \usepackage{booktabs}
 \usepackage{graphicx}
\usepackage{enumitem}
\usepackage{xspace}
\usepackage{xcolor}
\usepackage{colortbl}

\usepackage{hyperref}
\usepackage{cleveref}

\definecolor{blue}{HTML}{0066cc}
\definecolor{purple}{HTML}{660099}

\newcommand{\red}[1]{\textcolor{red}{#1}}

\newcommand{\blue}[1]{\textcolor{blue}{#1}}
\newcommand{\grey}[1]{\textcolor{gray}{#1}}
\newcommand{\orange}[1]{\textcolor{orange}{#1}}
\newcommand{\purple}[1]{\textcolor{purple}{#1}}

\newtheorem{thm}{Theorem}

\newtheorem{ex}{Example}

\newtheorem{example}[ex]{Example}
\newtheorem{definition}[thm]{Definition}

\newlength{\listingindent}                
\setlist{leftmargin=*,itemsep=0pt}

\DeclareMathAlphabet{\mathit}{T1}{cmr}{m}{it}

\newcommand{\stitle}[1]{\smallskip\noindent\textbf{#1}}

\newcommand{\sys}[0]{\texttt{dvl}\xspace}

\newcommand\vldbdoi{XX.XX/XXX.XX}
\newcommand\vldbpages{XXX-XXX}
\newcommand\vldbvolume{14}
\newcommand\vldbissue{1}
\newcommand\vldbyear{2020}
\newcommand\vldbauthors{\authors}
\newcommand\vldbtitle{\shorttitle} 
\newcommand\vldbavailabilityurl{URL_TO_YOUR_ARTIFACTS}
\newcommand\vldbpagestyle{plain}

\begin{document}
\title{A Formalism and Library for Database Visualization}

\author{Eugene Wu, Xiang Yu Tuang, Antonio Li, Vareesh Bainwala}
\affiliation{%
  \institution{Columbia University}
}
\email{ew2493,xt2280,asl2237,vb2589@columbia.edu}

%
%
%
%

%

\begin{abstract}
  Existing data visualization formalisms are restricted to single-table inputs, which makes existing visualization grammars like Vega-lite or ggplot2 tedious to use, have overly complex APIs, and unsound when visualization multi-table data.  This paper presents the first visualization formalism to support databases as input---in other words, {\it database} visualization.   A visualization specification is defined as a mapping from database constraints (e.g., schemas, types, foreign keys) to visual representations of those constraints, and we state that a visualization is {\it faithful} if it visually preserves the underlying database constraints.  This formalism explains how visualization designs are the result of implicit data modeling decisions.    We further develop a javascript library called \sys and use a series of case studies to show its expressiveness over specialized visualization systems and existing grammar-based languages.

\end{abstract}

\maketitle

\pagestyle{\vldbpagestyle}
\begingroup\small\noindent\raggedright\textbf{PVLDB Reference Format:}\\
\vldbauthors. \vldbtitle. PVLDB, \vldbvolume(\vldbissue): \vldbpages, \vldbyear.\\
\href{https://doi.org/\vldbdoi}{doi:\vldbdoi}
\endgroup
\begingroup
\renewcommand\thefootnote{}\footnote{\noindent
This work is licensed under the Creative Commons BY-NC-ND 4.0 International License. Visit \url{https://creativecommons.org/licenses/by-nc-nd/4.0/} to view a copy of this license. For any use beyond those covered by this license, obtain permission by emailing \href{mailto:info@vldb.org}{info@vldb.org}. Copyright is held by the owner/author(s). Publication rights licensed to the VLDB Endowment. \\
\raggedright Proceedings of the VLDB Endowment, Vol. \vldbvolume, No. \vldbissue\ %
ISSN 2150-8097. \\
\href{https://doi.org/\vldbdoi}{doi:\vldbdoi} \\
}\addtocounter{footnote}{-1}\endgroup

\ifdefempty{\vldbavailabilityurl}{}{
\vspace{.3cm}
\begingroup\small\noindent\raggedright\textbf{PVLDB Artifact Availability:}\\
The source code, data, and/or other artifacts have been made available at \url{\vldbavailabilityurl}.
\endgroup
}

\section{Introduction}
\label{s:intro}

Modern data analysis heavily relies on the ability to visualize and explore data.  The majority of visualization libraries, dashboards, and systems are built on top of graphical grammars, which provide a versatile and compositional method to compose data visualizations.  These grammars form the foundation of most data visualization systems including nViZn~\cite{Wilkinson2001nViZnA}, ggplot2~\cite{Wickham2010ALG}, Tableau~\cite{stolte2002polaris}, and vega-lite~\cite{Satyanarayan2018VegaLiteAG}, and their structured nature simplifies the development of advanced features such as visualization recommendation~\cite{mackinlay1986automating,mackinlay2007show,2019-draco}, visualization linters and consistency checks~\cite{Mcnutt2018LintingFV,qu2017keeping}, automated exploration~\cite{wongsuphasawat2015voyager}, and meta-analysis of visualizations~\cite{chen2023state,Mcnutt2018LintingFV,Mcnutt2022NoGT}.

\begin{figure}
    \centering
    \includegraphics[width=\columnwidth]{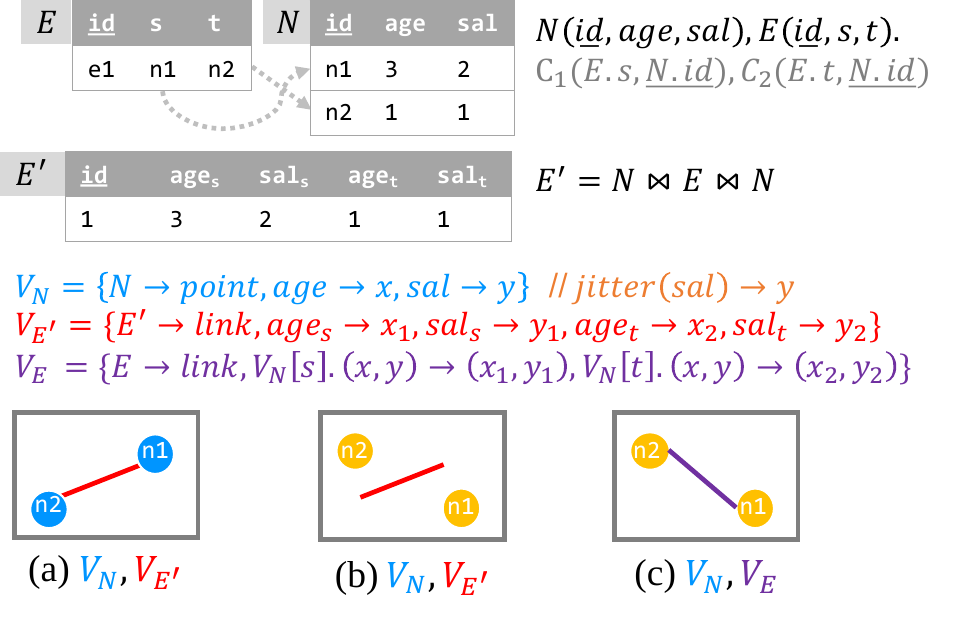}
    \caption{In the node link visualization, (a) the links \red{$V_{E'}$} only appear to connect the points \blue{$V_N$}. (b) The points and links become inconsistent if \orange{$V_N$} changes (e.g., adding \orange{jitter} to the y position). (c) \sys preserves the foreign key relationships by referring to the mark positions (e.g., \purple{$V_{N}[s].(x,y)$}) in the visual mapping \purple{$V_{E}$}.}
    \label{fig:nodelink}
\end{figure}

The theory that underlies graphical grammars is Bertin's data-to-visual (or pixel) mapping: tuples map to marks (e.g., points), and data attributes map to the mark's visual channels~\cite{Bertin1983TheSO}.  For instance, the points in \Cref{fig:nodelink}(a) visualize a table $N(id, age,sal)$ as a scatter plot \blue{$V_N$} where each row is mapped to a point mark, and $N.age$ and $T.sal$ are respectively mapped to the x and y position channels (ignore the red edges).
Wilkinson's Grammar of Graphics~\cite{wilkinson2012grammar} then extended this mapping with scales, mark types, facets, and other components to develop a {\it Graphical Grammar} that defines a design space of visual representations of a dataset.   This grammar-based formulation has been greatly successful in practice and theory.

However, Bertin's theory and its derivative grammars are predicated on a strong assumption: that the input is a single-table.   
This assumption has been codified in prominent models such as the Information Visualization Reference Model~\cite{card1999readings}, which define a pipeline that first transforms data {\it into a single table}, applies visual mappings, and then transforms the resulting view.  

Yet most information does not conform to a single table and is relational in nature: ranging from hierarchical JSON documents, to complex RDF graphs, to large multi-table relational databases. 
This gap means that visualizing data that is modeled as multiple tables using existing graphical grammars is tedious and unsound.  Further, these grammars only express a limited space of visualization designs and expose overly complex APIs.

\stitle{(L1) Tedious.} Current graphical grammars are tedious because the expectation is that the user is expected to ''prepare'' their data (extract, transform, filter, join)  and load it into the visualization tool, where each has a different set of APIs and expects a different schema or format.  There lacks a uniform abstraction that works with tables as they exist in the database.
For example, ggnetwork requires the user to load their data into a custom \texttt{network} object type and call the appropriate functions to load the edges and node attributes; D3's hierarchical algorithms expect a nested javascript object that represents the nodes and edges; Tableau expects users to join desired tables before they can be visualized, which can lead to semantic errors~\cite{Hyde2024MeasuresIS,Huang2023AggregationCE}.  

\stitle{(L2) Unsound.}
The need to fit complex relational data into a single table also makes existing graphical grammars theoretically unsound because, as this paper will show, they cannot guarantee that the visualization faithfully reflects the data. 
In fact, {\it faithfulness} is primarily qualitatively defined in the visualization literature~\cite{algebraicvis,hullmanconsistency} and this paper will formally define faithfullness and describe automatic ways to check faithfulness.
For instance, \Cref{fig:nodelink} connects the points in  \blue{$V_N$} based on the edges table $E(id,s,t)$.  In existing approaches, the user first computes $E'=N\Join E\Join N$, then maps $E'$ as links in \red{$V_{E'}$}.  However, the edges only {\it appear} to connect the points.  If the user jitters the points or applies a force-directed layout algorithm, then the positions of the points and edges become inconsistent because the visualization is unable to maintain their logical relationship (\Cref{fig:nodelink}(b)).   Our definition of faithfulness will show how this is a violation of referential integrity.

\stitle{(L3) Limited Expressiveness.} The single-table assumption also greatly reduces the space of visual designs that graphical grammars can express and makes it impossible for existing formalisms to express seemingly common designs such as node-link diagrams, parallel coordinates, tree diagrams, nested visualizations, nor tree maps.  
This leads users that wish to visualize relational data to resort to custom libraries for different ``data types''.    
For instance, a node-link visualization design that use e.g., a force-directed layout algorithm is not expressible in graphical grammars because it needs to reason about {\it two} tables.  As a consequence, the user needs to do more ``preparation'': run the layout to compute pixel coordinates of the nodes in a new table $N'=layout(N)$, derive a new edges table $E''=N'\Join E\Join N'$, and draw the points and links.  Unfortunately, this breaks the abstraction that the visualization maps data to pixels because $N'$ and $E''$ are tables of pixel coordinates and {\it not data}, so the grammar is merely used for rendering because all of the layout logic was applied during preparation.      

\stitle{(L4) Complex APIs.}
The limited expressiveness also increases the complexity of the visualization library due to the need to implement custom logic to support common functionality.  For instance, most graphical grammars support small multiples by introducing a custom x- and y-facet visual channel.  For instance, $\{N\to point, year\to x, price\to y, qtr\to fx\}$ renders a scatterplot comparing price and year for each quarter, and lays the subplots along the x-axis.   Internally, the library treats each quarter as a separate dataset with a single nominal attribute, draws a scatterplot where each quarter is rendered as a rectangle, and treats each rectangle as a new canvas to render a scatterplot for that quarter's data.      

While the first two steps are expressible in existing grammars, the third is not.  Thus all three steps are bundled within a custom facet component, and users are not able to express seemingly similar visualizations such as faceting by $cost$ and positioning subplots based on the $cost$'s quantitative (rather than nominal) value.  Further, the logic overlaps with other grammar functionality, such as view composition operations that horizontally and vertically organize views~\cite{Satyanarayan2018VegaLiteAG}, and can lead to confusion.



\smallskip
\subsection{Database Visualization}
There exists a simple and widespread data model that can express complex relational data---the relational data model~\cite{codd1970relational}---which is the basis of nearly all database systems today.   This model represents data as a {\it Database} that contains one or more tables along with constraints over attributes (e.g., data types, uniqueness) and tables (e.g., schemas, foreign key relationships).  

\begin{example}
    $N(\underline{id},age,sal)$ in \Cref{fig:nodelink} is a schema that specifies a primary key $\underline{id}$, and attributes $age$ and $sal$.  In addition, $C_1(E.s, \underline{N.id})$ specifies that $E.s$ is a foreign key that refers to the key $N.id$, and similarly for $C_2$.  These relationships are denoted using the dotted gray lines.
\end{example}

While databases can describe a wide range of real-world data, there lacks a theory to map a database to a visualization---a theory of {\it Database Visualization}.    This paper introduces such a theory: we model the visualization as a visual representation of the database contents (i.e., the rows and values) and define {\it faithfulness} as a visualization that preserves the database constraints.  
The key contribution is to treat visualizations as mappings of data constraints by distinguishing the definition of a visualization, which is derived from the schema (the constraints), from an instance of the visualization, which is populated using the contents of the database instance (the rows and values).   
In a sense, the visual representation of the constraints is the scaffold that users use to interpret and decode the visually encoded data values.   

Under this model, a visualization consists of four mappings: from each table's schema to a mark type, each attribute to a visual channel, each row to a mark, and each data constraint to either a corresponding constraint in the visualization or a visual encoding of the constraint.   Visual designs are a consequence of data modeling and transformation decisions or variations in the visual mapping specification.


\begin{example}
    \Cref{fig:nodelink}(c) and the \purple{purple definition of $V_{E}$} defines the node-link diagram as a database visualization.   $N$ is mapped to a set of point marks $V_N$, and $E$ is mapping to the set of link marks $V_E$ that preserves the foreign key constraints $C_1$ and $C_2$ in the visualization.  
    Rather than compute the links' x and y pixel positions, $V_{E}$ specifies their positions relative to the 
    start and end marks.  
    
    For instance, $V_N[s]$ can retrieve the node mark in $V_N$ that represents the source id $E.s$ by following the foreign key relationship $C_1(E.s,\underline{N.id})$ that uniquely identifies each mark in $V_N$.  We can render the starting position of the link for row $e_1$ by looking up the node with id $n_1$, and then retrieving the mark in $V_N$ labeled $n_1$.
    
    By doing so, the constraint $C_1$ is preserved in the point and mark definitions, and specifies that link positions are based on node positions and not vice versa.  Thus, even if the mark positions change (e.g., jitter), the visualization can ensure the links remain consistent.
\end{example}

The core formalism only requires three simple extensions to existing visual mappings.
The first is that marks such as $V_N$ are treated like tables, with well-defined schemas and constraints.
The second is the concept of {\it foreign attributes}, which allow a mapping specification for one table to reference values in a different table. For instance, \purple{$V_{E}$} refers to $V_N[s]$ in \Cref{fig:nodelink}(c).
The third are visual representations of foreign key constraints analogous to marks and channels that visually represent rows and attributes.  

We developed a Javascript library called \sys to support database visualizations.  The API for single-table visualizations is nearly identical to Observable Plot~\cite{plot} with a compact API to express foreign attributes and view nesting, as well an convenience functions for common patterns.  The prototype translates the specification into a task graph of SQL queries that run in a DBMS, and uses a thin layer to render the final data.   This makes it easy for users to write SQL to decompose and transform their database prior to visualization.

We showcase the benefits of database visualization in three ways.   
First, we show the correspondence between data modeling and visualization design by illustrating how four decompositions of a simple table $T(a,b)$ result in categorically different visualization designs that encompass categorical scatterplots, parallel coordinates, small multiples, and table visualizations.
Second, we introduce an extension to \sys to support space-filling layout algorithms and show how HIVE~\cite{hive}, a custom system for hierarchical space filling visualizations, can be trivially compiled and expressed to run on \sys.   
Third, we show how the space of visualizations expressible under the existing single-table formalism is just one option in a combinatorial space of designs, and is a direct consequence of the connection we establish between relational data modeling principles---such as normalization theory---and database visualization.   Using a case study, we show how our formalism expresses complex visualizations like ER diagrams, and enables a rich form of analysis that mixes SQL and complex visual forms under a simple abstraction. 


To summarize, this work contributes a new formalism for database visualization that connects  relational database theory and visualization.  We show that a small handful of primitives is sufficient to extend current single-table formalisms to support any relational database.  We describe a programmatic API called \sys, and show its utility for two use cases: translating a custom DSL into a handful of \sys statements, and expressing ER diagrams.
We end by discussing the theory's implications and sketching a database visualization research agenda.

\stitle{Scope.}
While important in building a complete system, we defer scalability and performance to followup work.  The focus of this paper is to present a new formalism that extends the theory of visualization from a single table to databases, illustrate its expressiveness and feasibility through examples and case studies.

\section{Background}

This section reviews relational data models and introduces a simple visual mapping formalism for single-table grammars. 

\subsection{Relational Data Modeling}

The purpose of a data model is to succinctly and correctly represent information using tables. The relational data model states that all data is stored as tables, along with a set of constraints that specify relationships between tables and uniqueness properties.  

\subsubsection{Tables and Databases}

$D=\{T_1,\dots,T_n,C_1,\dots,C_m\}$ is a database that contains tables $T_i$  and constraints $C_i$.
A table is defined by a schema that abstractly represents a type of entity and its structure, and a set of rows that describes the entities. 
A schema $T(a_1,\dots,a_n)$ specifies the table name $T$ and a list of attributes denoted $S_T=(a_1,\dots,a_n)$. An attribute $T.a_i$ has a type $T.\mathbb{D}_i$  that specifies its domain of valid values. Each row $r$ is a tuple of attribute values where $r.a_i\in T.\mathbb{D}_i$ is in its attribute domain.

A key is a set of attributes that is non-null and unique for every row in the table.   For convenience, we assume every table $T$ has a ``primary'' key $T.id$ that is used to distinguish two rows from each other and has no other semantic meaning.  Other attributes may also be denoted as keys (e.g., SSN).  We underline the primary key when listing a schema e.g., $T(\underline{id},a)$ has primary key $T.id$.    We will use the following shorthand notation: $T[x]$ denotes the row whose primary key has value $x$ or evaluates to true if $x$ is a predicate, $r.a$ accesses attribute $a$, and $r.(a,b)$ returns a tuple containing the values of $a$ and $b$.

\begin{example}\label{e:ex1}
    The schema $custs(\underline{id}, name, city, state)$ stores information about customers.
    $custs[1].(name,city)$ retrieves the name and city of the customer with id 1.
\end{example}

\subsubsection{Constraints}\label{sss:constraints}

Database constraints are boolean statements over the database that hold true irrespective of the table contents.  Developers rely on constraints to quickly understand what properties hold over the database, and database systems use them to automatically ensure data integrity.   We will show that, similarly, visualization users rely on visual representations of data constraints to make sense of the visualization.  In addition to schema constraints above, we will focus on two important classes, Keys and Foreign Keys, summarized in \Cref{tab:constraints}.  

\stitle{Attribute Type and Active Domain.}  As mentioned above, the type $T.\mathbb{D}_i$ is a constraint over the domain of valid values of the attribute $T.a_i$, and is potentially unbounded (e.g., $year$ has type int).  The attribute's {\it active domain}~\cite{lausen1998logical} is the subset of its domain that is actually present in the database (e.g., $year$ has values $2000-2024$).  Visualization scale functions (e.g., a linear scale that maps $year$ to the $x$ axis) typically map from the active domain to a value in the visual channel.   The active domain is visually depicted as axes for spatial channels or guides for other channels.

\stitle{Keys.} A key $\underline{T.X}$ is a subset of attributes $X\subseteq S_T$ that uniquely identify rows in the table; we drop the table reference if it is clear from context.  The primary key $\underline{id}$ is always a key, but other attributes may identify rows as well.    The benefits of keys will arise when visualizing foreign key constraints.  For instance, $\underline{custs.id}$ is the primary key in \Cref{e:ex1}.  Formally, it states:
\begin{align*}
    r_1.X=r_2.X \implies r_1=r_2\ \forall r_1, r_2\in T 
\end{align*}

\stitle{Foreign Keys.} A foreign key relationship $C(S.X,T.Y)$ specifies that a subset of attributes $S.X\subseteq S_S$ refer to attributes $T.Y\subseteq S_T$ of another table. Formally, it states that for each row in $S$, its values of $X$ are the same as the values of $Y$ for some row in $T$\footnote{Codd's definition in his seminal relational algebra paper~\cite{codd1970relational} is more strict, and  states that $S.X$ is not a key and $T.Y$ is a primary key.  We relax this definition in order to model relationships as used in practice.  }.  In other words, $S.X$ shares $T.Y$'s active domain:
\begin{align*}
    \forall s\in S \exists t\in T,\ s.X = t.Y
\end{align*}
Foreign keys are used to model three types of relationships.
\begin{itemize}[leftmargin=*]
    \item \stitle{1-1 relationship}, or one-to-one, when both $S.X$ and $T.Y$ are keys of their respective tables (e.g.,  $C(\underline{S.X},\underline{T.Y})$).  In such a case, the two tables effectively describe the same entity.   This is equivalent to $C(\underline{S.X},T.Y)\land  C(S.X,\underline{T.Y})$.
    \item \stitle{N-1 relationship}, or one-to-many, when $T.Y$ is a key and $S.X$ is not (e.g., $C(S.X,\underline{T.Y})$).  Each entity in $T$ may be referenced by multiple entities in $S$.  This is often used to denote hierarchical relationships and is the definition used in Codd's seminal work that introduced the relational data model~\cite{codd1970relational}.
    \item \stitle{N-M relationship}, or many-to-many, is modeled using an intermediate relation $W(X,Y)$ with constraints $C(W.X,\underline{S.X})$ and $C(W.Y,\underline{T.Y})$.  This is often used to represent social networks and other node-link graph structures where $S$ and $T$ represent nodes, and $W$ models the set of edges.
\end{itemize}

\begin{table}
    \centering
    \begin{tabular}{rll}
        \textbf{Name  }   & \textbf{Schema}   & \textbf{Constraints}\\
       Users     &  $U(\underline{id},username)$        &\\
       Settings  &  $S(\underline{uid}, passwd, \dots)$ & $C_1(\underline{S.uid},\underline{U.id})$\\
       Posts     &  $P(\underline{pid},\underline{uid},text)$ & $C_2(\underline{P.uid},\underline{U.id})$\\
       Follows   &  $F(\underline{uid},\underline{fid})$ & $C_3(\underline{F.uid},\underline{U.id}), C_4(\underline{F.fid},\underline{U.id})$
    \end{tabular}
    \caption{Constraints in a social network.  Users are identified by \underline{id}, 
  are 1-1 with their settings, and 1-N with their posts.  Users can follow and be followed by many users.  }
    \label{tab:constraints}
\end{table}


\begin{example}\label{e:ex2}
Continuing \Cref{e:ex1}, $items(\underline{id},name,descr,price)$ stores items that can be purchased and $orders(\underline{cid,iid},qty)$ tracks the quantity of items  customers have ordered, with
two constraints $C(orders.cid,\underline{custs.id})$ and $C(orders.iid,\underline{items.id})$.
\end{example}

\subsection{A Simple Single-Table Formalism}
We now introduce a simple formalism to describe graphical grammars that visually map an input table $T$ to an output view $V$ (a {\it visual mapping}). 
The notation focuses on data attributes, marks, mappings from data attributes to mark properties, and scales.  

A view $V$ is defined as a mapping from input table $T$ to a mark type, each row in $T$ to a mark that is indexed by the table's keys, and data attributes to the mark's properties (often called a visual channel).    
$V$ also has a spatial extent $V.ext=(x,y,w,h)$---the position and size of its canvas.
The notation $a\overset{s}{\to}v$ specifies an {\it aesthetic mapping} from data attribute $a$ to mark property $v$, that uses the scale function $s$. Scale functions can be named so that they may be referenced in multiple views; otherwise each mapping uses a different scale function.
For instance, the following creates a scatterplot that maps year to the mark's x property and price to y using linear scales, and the primary key $id$ using an identity function.  Its spatial extent is positioned at the origin and $800\times 600$ pixels in size.
\begin{align*}
     V &= \{ T\to point,\ T.id\overset{s_{id}}{\to}id,\ T.year\overset{s_x}{\to} x,\ T.price\overset{s_y}{\to}y\\
     &\hspace{1.5em}| (0,0,800,600)\to ext \}\\
     s_{id} &= identity\hspace{2em}
    s_x = linear\hspace{2em}
    s_y = linear 
\end{align*}
This formulation makes clear that $V$ is in fact a  table of marks with schema $V(type,\underline{id},x,y,\dots)$ that contains its mark type, keys, and mark properties as attributes. As a table, $V$ also has a 1-1 relationship with $T$ using the constraint $C(\underline{V.id},\underline{T.id})$.  Transitively, $V$ preserves the same relationships  that $T$ has with other tables. 

Each mark $m\in V$ also has a spatial extent $m.ext$. To keep the notation simple, we will omit the primary key mappings because they are always present, the table name if it is evident, and scale functions and spatial extent if they are not important to an example.    For instance, the following is equivalent to the above example:
\begin{align*}
    V &= \{ T\to point, year\to x, price\to y \}
\end{align*}

\section{Database Visualization}\label{s:model}

We wish to extend the single-table formalism to map a multi-table database $D$ to a multi-view visualization $P$ ($P$ for plot) that consists of one or more views.  
The main principle is to define visualization mappings over the database constraints, whereas the contents of the database instance simple ``fill in'' the marks and channels of the visualization.
To motivate this formalism, we first re-frame single-table grammars as visually mapping constraints.

We then show that only two new concepts are needed to express database visualization.  The first are {\it foreign attributes} so the visual mapping for a table  can reference attributes in other (foreign) tables (\Cref{ss:joins}).  This makes it possible to render foreign key relationships as marks, and for the visualization to maintain consistency between marks even as their attributes change.  

The second are visual representations of foreign key relationships. 
In the same way that there are many visualization designs and mark types suitable for different table schemas, foreign key constraints can be represented by there are a range of visual designs to express foreign key constraints.   We present two initial classes of designs: those that explicitly render relationships as marks, and those that implicitly denote relationships using spatial layouts.



Our focus here is to present a formalism under the assumption that the user has performed all desired data modeling, transformation and filtering operations in the database, and has identified the subset of tables and attributes to visualize.  In other words, the user wishes to visualize all data in the prepared database.   
The subsequent sections will introduce a programmatic API based on this formalism, syntactic sugar and heuristics to infer sensible defaults, and examples of how the formalism composes with data modeling and transformation operations to express a wide range of visualization designs.  



\subsection{Graphical Grammars Map Constraints}\label{ss:mapkeys}

In the relational model, the structure (or schema) of the database corresponds to the set of table schemas and constraints over them, while the contents of the database correspond to the rows and their attribute values.  As grammar-based visualization is a mapping from data to pixel space, we argue that the structure of the visualization---{\it how} marks, mark properties, axes, subplots, and views are laid out---is defined by mapping data constraints, while its contents---the set of marks and their properties---are defined by the contents of the database. 
We will elaborate by reframing single-table visualization under this model, and then show how its principles extend to multi-table database visualization.

Let us revisit the specification that renders $T(\underline{id},a,b,c)$ as a scatterplot, but now from the perspective of its constraints:
$$M = \{T\to point, a\to x, b\to y, c\to color\}$$
The table $T$ has two sets of constraints: domain constraints for each attribute and key constraints.

\stitle{Attribute Domains.}
The attribute domains of $a$ and $b$ determine the spatial layout of the marks, and $c$'s domain determines the mark colors.  Spatial axes and guides are explicit visual representations of these attribute domains.

\stitle{Key Constraints.} 
Key constraints distinguish rows from each other, which requires a way to distinguish them in the visualization---namely by creating a separate mark for each row.  
Naively, the constraint $\underline{T.id}$ is satisfied because $\underline{M.id}$ is always copied and is a key.   However, this is not sufficient because the visualization is not a set or bag of rows---its marks are spatially positioned in a 2D space.  Thus, different marks may appear indistinguishable and violate the constraint.    This is commonly called ``overplotting''.
\begin{align*}
    indistinguishable(m_1,m_2)\implies m_1=m_2\hspace{3em}m_1,m_2\in M
\end{align*}
Unfortunately, the way to detect indistinguishability is rooted in perceptual and cognitive questions beyond the scope of this work, and rely on the designer's judgement to adequately address.   However, we can categorize the common ways to address this constraint violation into four methods:
\begin{itemize}[leftmargin=*]
    \item \stitle{Change the Data.}  The field of data cleaning addresses constraint violations by modifying the data---deduplication, value imputation, extraction---so that it conforms to the user's desired form.    However, there are additional options in visualization.
    \item \stitle{Change the Constraints.} Designers often resolve overplotting by grouping by the spatial attributes (e.g., $T.(year,price)$) to render a heatmap.  This changes the data constraints by making the spatial attributes the key.   A similar operation is to reduce the opacity, which groups by individual pixels and makes each pixel the key.    
    \item \stitle{Change the Mapping.}  The designer can instead map $id\to x$ and $year\to color$ to remove overlaps.   
    \item \stitle{Change the Layout.}  The designer may perturb the mark positions by e.g., jittering the points to reduce overlaps.  
\end{itemize}
Irrespective of how a constraint violation is resolved, if the constraint is satisfied, it implies that the spatial extent, meaning the data attributes (e.g., $a,b$) mapped to the spatial attributes in this paper, must also form a key.
Further, since the purpose of a key constraint is to distinguish data rows, satisfying any of the table's key constraints in the visualization is equivalent to satisfying all of them.   This gives the visualization author flexibility in choosing {\it which} key to map to the visualization.

%

\begin{table}
    \centering
    \begin{tabular}{rl}
        \textbf{Data Constraint} & \textbf{Visual Representation} \\
        Key & \red{Mark per row} \\
        Attribute Domain& \blue{Visual channel, axis/guide} \\
        1-1 Relationship& \red{Mark}, \blue{share scale, align, merge marks}  \\
        1-N Relationship& \red{Mark}, \blue{nest} \\
        N-M Relationship& \red{Mark} \\
    \end{tabular}
    \caption{Summary of visual representations for different database constraints.  \red{Red text} and \blue{blue text} denote explicit and implict representations, respectively.}
    \label{tab:my_label}
\end{table}

%

\subsection{Visual Mappings with Foreign Attributes}\label{ss:joins}

Multi-table databases and foreign key constraints make it possible and useful to reference multiple tables in a visual mapping.  
Formally, we extend a visual mapping to map {\it expressions over data attributes} to mark properties, and define a special class of functions $f(S.X)=T.W$ that takes as input data attributes $S.X$ and outputs data attributes $T.W$, where $S.X\leadsto T.W$ is a {\it valid path} composed of $N-1$ foreign key relationships.  This ensures that the reference to $T.W$ unambiguously specifies a single $T$ row's attribute values i.e., for any row in $S$, it corresponds to exactly one row in $T$.   If $S.X\leadsto T.W$ is composed of $1-1$ relationships, then $f(S.X)$ is also a key of $S$.   Let us examine some examples of $f()$ in action.

\begin{example}\label{ex:scatterfkey}
$V$ renders $S$ as a scatterplot, but maps attributes $f_1(S.c)=T[c].a$ to $y$  and $f_2(S.c)=T[c].b=T.b$ to $color$.   This is allowed because $S.c\leadsto T.id$ is a $N-1$ relationship.  $T[c].b$ and $T.b$ are equivalent because the reference is unambiguous.
\begin{align*}
    &T(\underline{id}, a, b), S(\underline{id}, c, d), C_1(\underline{T.id}, S.c)\\
    V &= \{S\to point, d\to x, T[c].a\to y, T.b\to color\}
\end{align*}

\end{example}

\noindent If the path is unique, $T$ can be directly referenced (e.g., $T.a$); otherwise, a fully qualified path to $T$'s attribute is needed (e.g., $T[S.c].a$).   For instance, the definition of $V_E$ in \Cref{ex:nodelinkfkey} cannot use $V_N.(x,y)$ because the edges table $E$ has two foreign key references to $N$ through $E.s$  and $E.t$.  Thus, it needs to specify $V_N[s]$ or $V_N[t]$.

A powerful consequence of foreign attributes is that {\it any} table with a valid path from $S$ is a valid reference---including mark tables, which have a 1-1 relationship with their data table. 
Further, if the foreign attribute is a mark property (e.g., $V_N[t].(x,y)$ in \Cref{ex:nodelinkfkey}), its values are already in pixel-space so scales are not necessary.

\begin{example}\label{ex:nodelinkfkey}
   The visualization in \Cref{fig:nodelink} renders edges $E$ by referencing the point marks $V_N$ for the nodes table $N$.  The definition of $V_E$ can reference $V_N$ because $E\to N\to V_N$ is a valid path:
\begin{align*}
    V_N &= \{N\to point, age\overset{sx}{\to} x, sal\overset{sy}{\to} y\}\\
    \purple{V_E}&= \purple{\{E\to link, V_N[s].(x,y)\to (x1,y1), V_N[t].(x,y)\to (x2,y2) \}} \label{eq:join}
\end{align*}
\end{example}

\begin{figure}[h]
    \centering
    \includegraphics[width=.9\columnwidth]{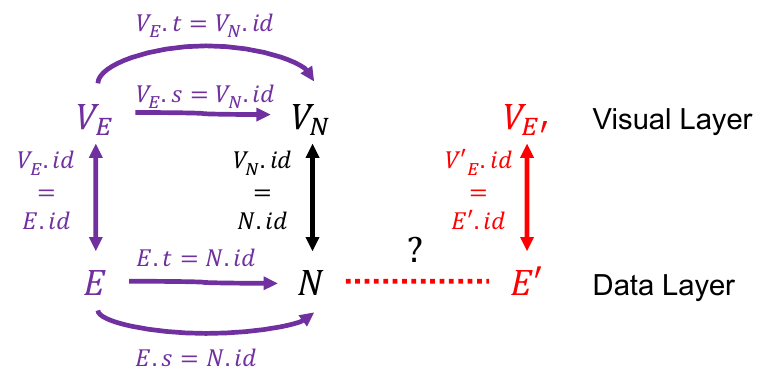}
    \caption{Using foreign attributes (\purple{in purple}) preserves foreign key relationships between the nodes and edges tables {\it and between their views $V_E$ and $V_N$}.   The single-table approach (\red{in red}) does not preserve the relationship between $V_N$ and $V_{E'}$, and may break the relationship between $N$ and $E$ as well.}
    \label{fig:foreignattrgraph}
\end{figure}

The \purple{purple text} in the example matches the \purple{purple portion} of \Cref{fig:foreignattrgraph}, and illustrates the foreign key constraints between the base tables $N$, $E$ and the node and edge views $V_N$ and $V_E$, respectively.   For instance, the two directed edges $E\to N$ denote the two relationships $C_1$ and $C_2$, where an arrow head denotes a key.   Similarly, there are 1-1 relationships $E\leftrightarrow V_E$ and $N\leftrightarrow V_N$.  Finally, $V_E$ is defined with references to $V_N$, which define the two arrows $V_E\to V_N$.   In contrast to $V_{E'}$ in \Cref{fig:nodelink}, the foreign key references between the data tables are preserved in the visualization

\begin{figure}[h]
    \centering
    \includegraphics[width=\columnwidth]{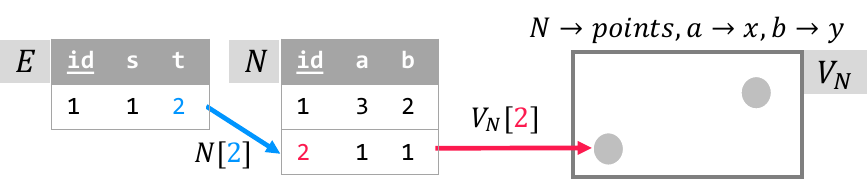}
    \caption{The foreign attribute expression $V_N[E.t].(x,y)$ follows foreign key relationships to index into $N[\blue{2}]$ using $\blue{E.t}$, index into $V_N[\red{2}]$ using $\red{N.id}$, then retrieve the $x,y$ mark properties.  }
    \label{fig:enter-label}
\end{figure}

\stitle{Comparison to Single-table Grammars.}
As mentioned in \Cref{fig:nodelink}, the above example can only be superficially mimicked using single-table grammars.   To do so, the developer can modify the edges table to instead store the $age$ and $sal$ of the source and target nodes and then render them using the same scales:
\begin{align*}
    &\dots,\red{E'(\underline{id},age1,sal1,age2,sal2)},\dots\\
    \red{V_{E'} =}&\red{\{E'\to link, (age1,age2)\overset{sx}{\to}(x1,x2), (sal1,sal2)\overset{sy}{\to}(y1,y2)\}}
\end{align*}
Although syntactically similar, there are two drawbacks to the single-table approach.  
First, the edges table $E$ has now lost its relationship between $N$---for a given $age1,sal1$ pair, we cannot determine its corresponding node (dashed edge in \Cref{fig:foreignattrgraph}).   Although this could be remedied by preserving $s$ and $t$, the second drawback is more fundamental: $V_E'$ only {\it happens} to be rendered in a way where the links start and end at the appropriate node marks.   If $V_N$ were updated to e.g., map $hobby$ to the x-axis or to use a force-directed layout algorithm, the two views would become inconsistent because the edges in $V_E$ would not be correctly updated.  This is because there are no relationships between $V_N$ and $V_{E'}$ that can be enforced by the visualization.  In contrast, using foreign attributes allows the visualization to automatically preserve the foreign key constraints between $E$ and $N$ in the visualization between $V_N$ and $V_E$.

\subsubsection{Foreign Attribute Aggregation}
The foreign attribute formalism assumes that the referenced table is accessed using a key (e.g., $V_N[s]$ in \Cref{ex:nodelinkfkey}) so that it returns a unique row.  But in principle, a table can be accessed using any filter condition, and as long as the attribute of the matching rows are aggregated into a scalar, the expression is unambiguous.    For these reasons, any expression that references a foreign attribute is valid as long as it is guaranteed to return a single scalar, however aggregations do not preserve foreign key constraints.  For instance, $max(V_N[color='green'].x)$ would return the largest x-position for green marks.

\subsection{Mapping Foreign-key Constraints}

Foreign key constraints are used extensively to model network and hierarchical data.  For instance, the former is modeled as node and edges tables as in \Cref{fig:foreignattrgraph}, while the latter is modeled as a single nodes table $N(\underline{id}, parent, \dots)$ with a constraint $C(N.parent,\underline{N.id})$.  We thus draw on existing visual metaphors from network~\cite{gibson2013survey}, hierarchical~\cite{schulz2010design,bruls2000squarified} (e.g., treemaps), and multi-view
visualizations to propose methods to visually map a foreign key constraint $C(S.X, T.Y)$.  

We categorize designs as {\it explicit} if they render marks to encode the relationship or {\it implicit} otherwise.
For instance, network visualizations render each node as a mark and the relationship between two nodes using an explicit link; the spatial placement of the node marks is based on various layout algorithms.   In contrast, treemaps use spatial containment to render parent-child relationships by subdividing a parent's spatial extent to render its children.  Finally, multi-view visualizations use shared scales and alignment to denote common attributes across multiple views.

The same way a data attribute can be encoded in more than one visual channels,    a relationship may be expressed using a combination of explicit and implicit methods.  Furthermore, just as new visualization designs continue to be discovered, we do not claim the designs below are exhaustive; they simply represent a set of simple designs.  Below, we will assume that $S$ and $T$ each have a single mark table $V_S$ and $V_T$.

\subsubsection{Explicit Representation: Marks.}

What does it mean to visualize a constraint?
Tables are populated with rows to instantiate marks and values to set  mark properties, 
but a foreign key constraint $C(S.X,T.Y)$ is an abstract logical statement about data.

A naive approach might instantiate $C$ as a table $T_C(X,Y)$ and then render it as a single-table visualization.  Although syntactically correct, it is semantically nonsensical because $S.X=T.Y$ for every row in $T_C$.    In addition, $S$ and $T$ are unlikely to map keys $S.X$ and $T.Y$ to visual channels, so the user cannot visually identify the correspondence between $V_{T_C}$ and $V_T$ or $V_S$.  This violates the foreign key relationship.


Foreign attributes (\Cref{ss:joins}) are a powerful way to enforce foreign key constraints in the visualization by simply referencing them in the visualization mapping.     
The most direct way is to treat $C$ as a table in the naive approach but map functions $f_1(S.X)$ and $f_2(T.Y)$ to mark keys.

\begin{example}
For the following hierarchical dataset, we render nodes as points ($V_N$) and links from child to parent nodes ($V_{E_1}$) because the spatial position of the point mark is a key (\Cref{ss:mapkeys}).  The specific positions of the points do not matter, so its specification is denoted with ``$\dots$''.
\begin{align*}
      &N(\underline{id}, p),  C(N.p, \underline{N.id})\\
V_N = &\{ N\to point, \dots \}\\
V_{E_1} = &\{ C\to link, V_N[id].(x,y)\to (x1,y1), V_N[p].(x,y)\to (x2,y2) \}
\end{align*}
\end{example}
\noindent It is not necessary to render the constraint as a line between marks.  It is sufficient for $f_1(S.X)$ and $f_2(T.Y)$ to return keys (e.g., $N.id$, $V_N.(x,y)$) that are 1) mapped to the same mark's properties, and 2) those keys are present as data attributes or channels in $V_S$ and $V_T$, respectively.   
For instance, the following renders the foreign key constraint using a label instead:
%
%
\begin{example}\label{ex:explicit-labels}
Table $T$ contains a set of entities, $S$ contains the corresponding names of the entities, and there is a 1-1 relationship between $T$ and $S$.
We wish to render $T$ as points, and place each point's name 10 pixels to its right:
\begin{align*}
    &T(\underline{id}, a, b), S(\underline{id}, name), C(\underline{S.id}, \underline{T.id})\\
    V_T =& \{T\to point, a\to x, b\to y \}\\
    V_S =& \{S\to text, name\to text, 10\to dx, V_T[c].(x,y)\to (x,y)\}
\end{align*}    
Notice that $V_T[c].(x,y)$ is a key of both $S$ and $T$ because $V_T.(x,y)$ is a key and there are 1-1 relationships between $V_T$, $T$, and $S$.   
\end{example}

\begin{figure}
    \centering
    \includegraphics[width=0.5\columnwidth]{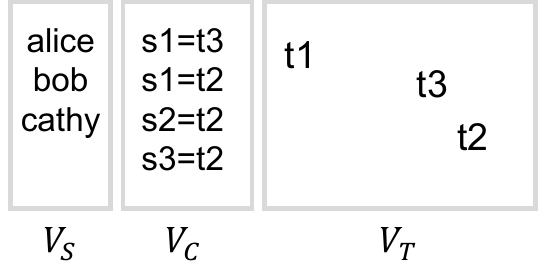}
    \caption{Constraint is explicitly represented as text that equates ids from $S$ and $T$.  The user can identify the corresponding $T$ mark by finding the point with the label, and the $S$ mark by its position along the y axis. }
    \label{fig:label}
\end{figure}

%
%
%
\noindent Since the main requirement is that the relationship's keys are mapped to properties of the same mark, the specific channel they are mapped to does not technically matter (though some channels may be more effective than others).     The following is an example of a less effective visual representation that still preserves the database constraints.

\begin{example}
\Cref{fig:label} renders the same dataset as \Cref{ex:explicit-labels}, but where $V_T$ renders labels of $T.id$ as a scatterplot and $V_S$ renders a list of names positioned by $S.id$.   $V_C$ renders a list of strings: each string is formatted to render the ids of the related entities;  we are treating $C$ as a table so $C.id\to y$ refers to $id^{th}$ relationship:
\begin{align*}
    V_T =& \{ T\to label, id\to text, a\to x, b\to y\}\\
    V_S =& \{ S\to label, name\to text, 0\to x, id\to y\}\\
    V_C = &\{ C\to text, 
    f"\texttt{\{}S.id\texttt{\}=\{}T.id\texttt{\}}"\to text,
    id\to y, `5em'\to x \}
\end{align*}    
\end{example}


\subsubsection{Implicit Representation: Spatial Nesting}

The major drawback of the explicit representation is that marks occupy the view's limited spatial extent and can lead to crowding or overlapping marks, which then violate the key constraints.  For this reason, an alternative is to use spatial organization to implicitly denote relationships.   We start with spatial nesting, which is used to denote N-1 relationships common in hierarchical data, exemplified by small-multiples, facetting, treemaps, and other nested designs~\cite{hive,bruls2000squarified}.

The key abstraction is that the parent entities in the relationship (the 1 side of the N-1 relationship) are rendered as marks whose spatial extents that can be the extents of the views that their child entities are rendered within.

\begin{example}
The following database contains annual price information for stores ($S$) and their latitude-longitude locations ($L$).  There is a N-1 relationship between $S$ and $L$.   $V_L$ renders $50\times50$ rectangles at each store location, and $V_S$ renders the price data as a scatterplot within its corresponding store location's rectangle.     This is specified using the foreign attribute mapping $V_L[cid].ext\to ext$.
\begin{align*}
    &L(id, lat, lon), S(id, cid, year, price), C(S.cid, \underline{L.id})\\
    V_{L} =& \{L\to rect, lat\to x, lon\to y, 50\to w, 50\to h\} \\
    V_{S} =& \{S\to point, year\to x, price\to y\ |\ V_L[cid].ext\to ext \} \\
\end{align*}
\end{example}

%

\noindent For convenience, we make this explicit using the $nest()$  operation, which lets the nested view be written independently of its parent.  The constraint $C$ can be omitted if it is unambiguous:
\begin{align*}
    V_L =& \cdots, \hspace{1em}
    V_S = \{S\to point, year\to x, price\to y \}\\
    & nest(V_S\ in\ V_L\ using\ C)
\end{align*}

\noindent The nesting specifies that only the $S$ rows for a given store should be rendered within the extent of its related mark in $V_L$.
From this perspective, a multi-view visualization is simply an application of spatial nesting.  There is a root table with one row whose extent is the entire canvas and all other tables have a N-1 relationship with the root table and thus spatially nested within its mark.  

\subsubsection{Implicit Representation: Shared Scales \& Alignment}

Nesting is  less suitable for N-M or 1-1 relationships.  In these cases, a common alternative is to share and possibly also align axes.  Returning to the constraint $C(S.X, T.Y)$, let the two tables be mapped to marks as follows, where a function of each table's key is mapped to some spatial channel $c$:
\begin{align*}
    V_S = \{ S\to \dots, f_S(X) \overset{s_S}{\to} c_S\}\hspace{2em}
    V_T = \{ T\to \dots, f_T(Y) \overset{s_T}{\to} c_T\}
\end{align*}
Recall from \Cref{sss:constraints} that a foreign key simply means $S.X$ and $T.Y$ share their attribute domains.  Thus, it suffices that the scaling functions' domains are equal: $s_S.domain \equiv s_T.domain$. 

\begin{example}
    \Cref{fig:label} uses shared scales to ensure the visualization is faithful.   Specifically, the expression \texttt{"{s.id}={T.id}"} is the cross-product, and thus shares, the domains of $T.id$ and $S.id$.
\end{example}

\begin{figure}
    \centering
    \includegraphics[width=.75\columnwidth]{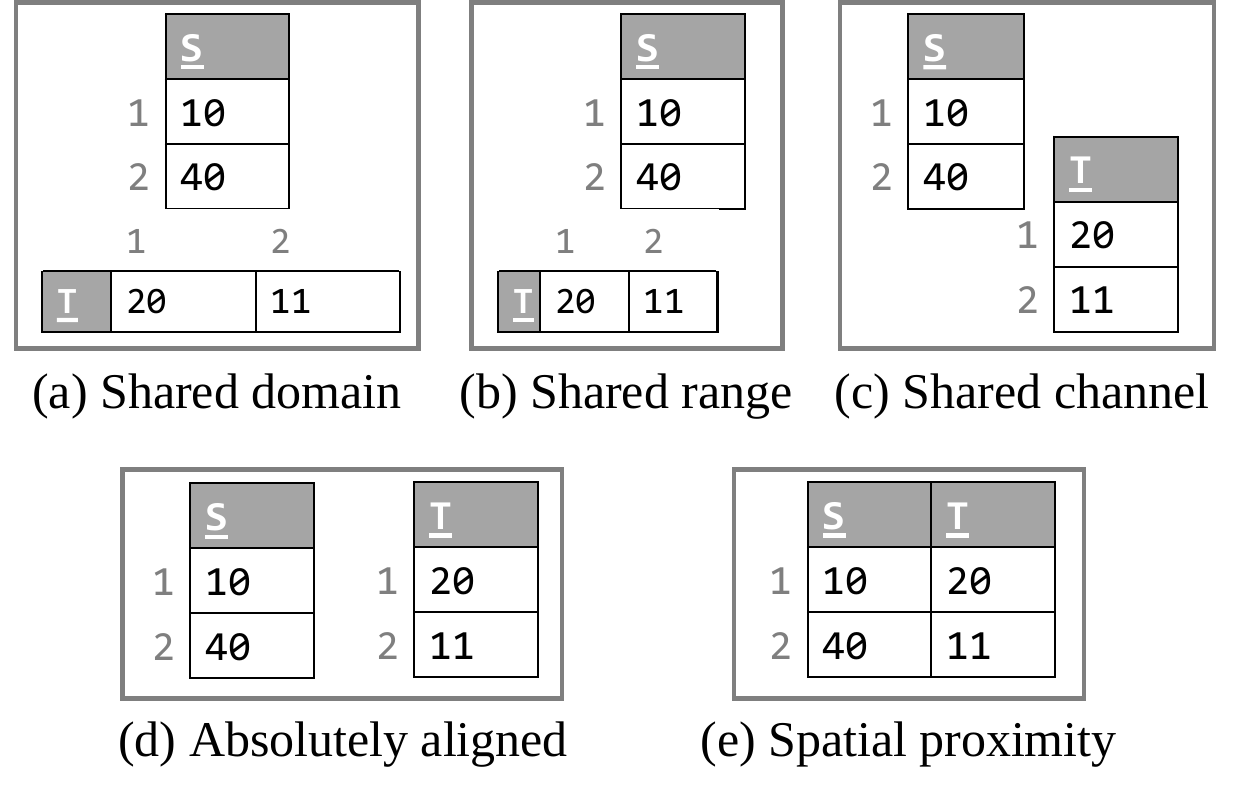}
    \caption{Foreign key relationship between $S$ and $T$ is preserved by (a) shared scale domains and can be progressively reinforced by (b) shared scale ranges, (c) shared channels, (d) absolute alignment of the views, and (e) spatial proximity.  }
    \label{fig:implicit-align}
\end{figure}

\begin{example}\label{ex:alignment}
    
\Cref{fig:implicit-align}(a) renders $S$ and $T$ as labels.  We 
include the ids in \grey{grey} and label each table for the reader's reference only.
The scales that map $S.id$ and $T.id$ have the same domain, so the rows are consistently ordered.   While it is technically possible to relate the $i^{th}$ rows from $S$ and $T$, it is very difficult.
\begin{align*}
    & T(\underline{id}, t), S(\underline{id}, s), C(\underline{S.id}, \underline{T.id})\\
V_S =& \{S\to label, id\overset{s_S}{\to} y, 0\to x, t\to text  \}\\
V_T =& \{T\to label, 0\to y, id\overset{s_T}{\to} x, s\to text  \}\\
&s_S.domain = s_T.domain
\end{align*}
\end{example}
\noindent In practice, these relationships are reinforced with redundant encodings.    \Cref{ex:alignment} illustrates common methods that build on each other (\Cref{fig:implicit-align}(b-e)):
\begin{enumerate}
    \item \stitle{Relative Alignment} enforces that the scale functions' ranges are also equal: $s_S.range = s_T.range$. This consistently order the marks so they are relatively positioned and sized in both views    
    \item \stitle{Shared Channel} requires that the spatial channels are the same e.g., $S.id$ and $T.id$ are both mapped to $y$.
    \item \stitle{Absolute Alignment} ensures that the spatial extents of the views $V_S$ and $V_T$ are aligned along the shared channel (e.g., $y$).   This alignment is recursively applied to views that $V_S$ and $V_T$ may be nested within. 
    \item \stitle{Spatial Proximity} ensures that the shared channels are positioned near each other to reduce separation effects~\cite{talbot2014four}.  When all four methods are applied, this results in a table visualization.  
\end{enumerate}
%


\subsection{Faithful Database Visualization}

Our formalism defines the primitives to map a prepared database to a visualization, where the designer has flexibility to prepare the database by transforming, decomposing, copying, joining, and dropping tables and constraints.
Allowing multiple input tables simplifies the structure of the mapping:
1) each table maps to one view; 
in a given view, 2) each row maps to one mark and 3) each attribute maps to a mark property; 
and 4) each constraint is {\bf preserved}.

For individual tables, (1-3) are satisfied in single-table formalisms, while only key constraints are applicable for (4).  We heuristically check key constraints by detecting whether there is more than 95\% area overlap between marks.
A foreign key relationship can be {\bf preserved} in several ways.
The first is if the foreign key is traversed as part of a foreign attribute expression that returns a key of the referenced table.  For example, in \Cref{ex:nodelinkfkey}, the visual mapping for $E$ maps the expression $V_N[s].(x,y)$ to the starting position of the link.  Since $(x,y)$ is a key of $N$\footnote{$(x,y)$ is a key of $V_N$, and there is a one-to-one mapping between $V_N$ and $N$.} this foreign reference preserves the foreign key constraint $C_1(E.s,\underline{N.id})$.
The second is if the constraint is treated as a table and explicitly mapped to a mark,
and the third is if it is preserved implicitly via nesting or shared scales.  

\begin{definition}[Faithful Database Visualization]
We call a database visualization {\it Faithful} if:
\begin{itemize}
    \item Every table $T$ is mapped to a view $V_T$,
    \item For each table, $\ge1$ key constraints are satisfied,
    \item For every foreign key constraint between tables $T$ and $S$, the views $V_T$ and $V_S$ preserve the constraint.
\end{itemize}
\end{definition}
\noindent This definition delineates a strict boundary between preparing the database---creating, modifying, dropping, or (de)normalizing tables---and visualizing the database.  For instance, if the same table should be rendered in $K$ views, the database should logically contain $K$ copies of the table (in practice, these copies are created implicitly and not materialized).   This reduces the logic for visual mappings, allowing more responsibility to be pushed to an underlying DBMS.


\begin{figure*}
    \centering
    \includegraphics[width=.85\textwidth]{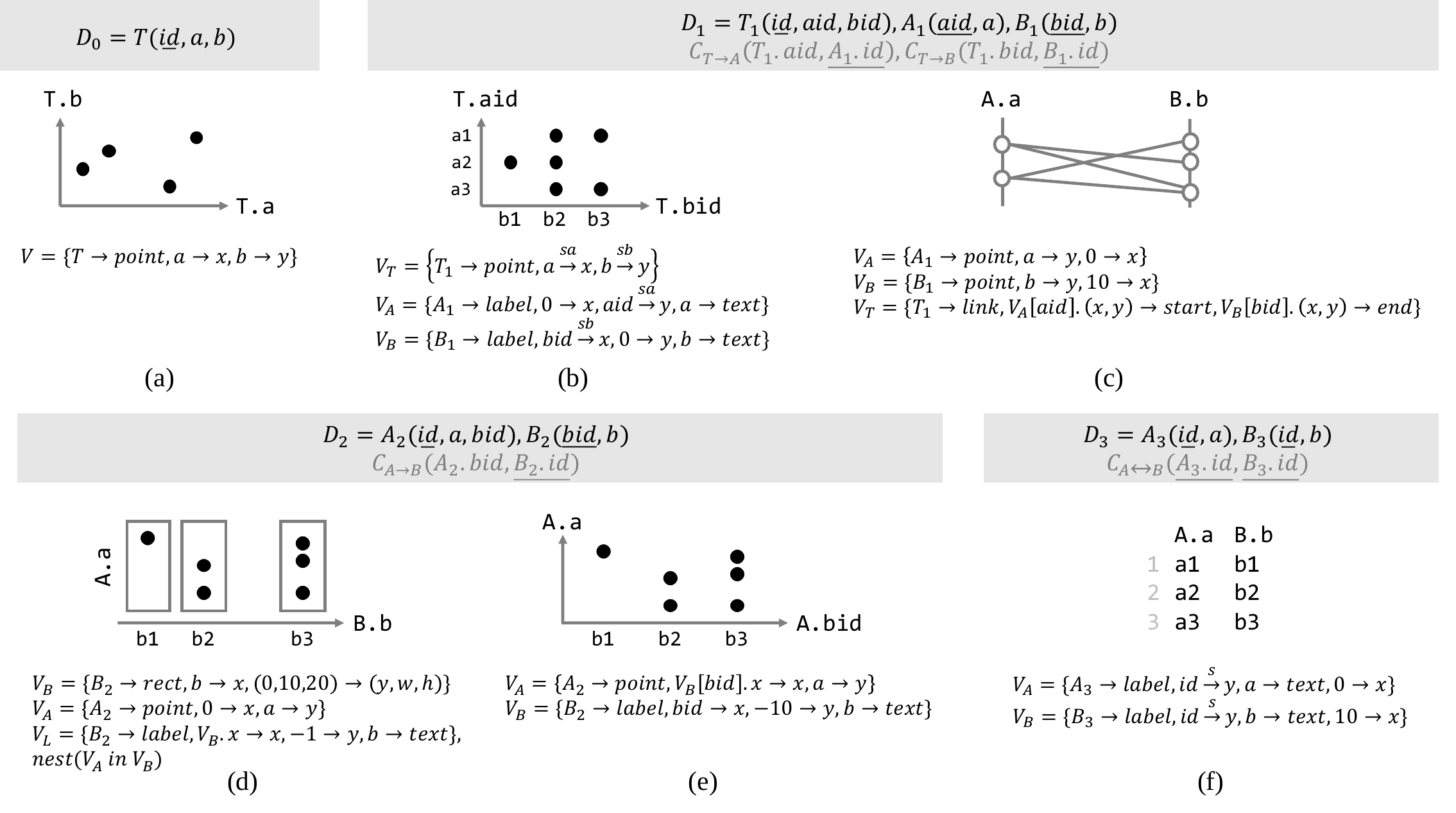}
    \caption{Examples that show how different visualization structures are needed to faithfully express different sets of tables and constraints. Common visualization designs are a consequence of data modeling decisions. }
    \label{fig:gallery}
\end{figure*}

\section{Database Visualization In Action}\label{s:impl}

We have implemented the \sys \texttt{\bf D}atabase \texttt{\bf V}isualization \texttt{\bf L}ibrary in Javascript.   It closely follows the familiar Observable Plot's API~\cite{observable} for single-table visualizations and introduces extensions to express the formalism in \Cref{s:model}.  Input data is stored, decomposed, and transformed in a database (\sys currently uses client-side DuckDB-wasm~\cite{kohn2022duckdb}).   Concretely, the library compiles the specification into a DAG of SQL queries for data transformations and javascript function calls for layout and rendering.  
\sys can (de)serialize specifications to JSON for portability.

This section will introduce \sys by recreating the visualization designs in \Cref{fig:gallery}.  The figure shows how different decompositions of the same table can lead to different visualization designs.  
In each example, we assume  the designer has initialized a new Plot \texttt{p}, and has loaded the appropriate tables in the database \texttt{db}:
{\small\begin{verbatim}  p = dvl.plot({width: 800, height: 600}) \end{verbatim}}

\subsection{Single-table API}

\sys follows Observable Plot syntax and internally calls it for rendering. The following\footnote{\texttt{dot} is often synonymous with \texttt{point} in visualization libraries.} renders \Cref{fig:gallery}(a).  The table $T$ may be passed in by name or as a Table object (e.g., \texttt{T = db.table("T")}).
{\small\begin{verbatim}  V = p.dot(T, { x: 'a', y: 'b'}) \end{verbatim}}
\noindent $D_1$ normalizes the attributes $a$ and $b$ as their own entities.  This is useful if the original $T$ encodes a many-to-many relationship between tables $A$ and $B$, as is common for graph data.   Below, the foreign key references are preserved due to shared scales.
\begin{example}[Punchcard]
    \Cref{fig:gallery}(b) renders the cross product $T.aid\times T.bid$ as a scatterplot $V_T$. $A$ and $B$ are rendered as labels along the y (in $V_A$) and x-axes (in $V_B$); the relationships are preserved by sharing the scales so that the labels are aligned with their positions in $V_T$. For instance, $V_A$ maps $aid\to y$ using the same scale $sa$ as in $V_T$. 
    \texttt{apply(scale, attr)} calls the \texttt{scale} during the render process but also tracks this metadata when checking faithfulness.   This visualization is often called a punchcard design.
     {\small\begin{verbatim}  sa = p.linear("T.a"), sb = p.linear("T.b")
  VT = p.dot(T, { x: apply(sb,'bid'), y: apply(sb,'aid')})
  VA = p.label(T, {x: 0, y: apply(sa, 'aid'), text: 'a'})
  VB = p.label(T, {x: apply(sb, 'bid'), y: 0, text: 'b'}) \end{verbatim}}
    
\end{example}

\subsection{Multi-table APIs}
\sys uses a small number of primitives to support multi-table APIs.

\subsubsection{Foreign Attributes}

A view is a type of table and can be referenced to retrieve its marks and their properties.   The primary method \texttt{V.get(filter, props?, cb?)} takes as input a filter condition, and optionally desired mark properties and a callback---essentially a selection, optional projection, and optional table function that returns a scalar. The filter is either a set of attributes in the referring table that act as search keys in \texttt{V}, a predicate, or a boolean function that takes a row as input.  \texttt{props} may be specified as a list of attribute names to be projected, or an object that renames the value to the key.  If not specified, it retrieves all mark properties.

 The callback \texttt{cb(markrows, datarow)$\to$scalar} takes as input the projected marks matching the filter (\texttt{markrows}) and the data row in the view specification's input data table (\texttt{datarow}), and returns a scalar.  This is intended to aggregate rows when filtering on non-keys so that \texttt{get()} is unambiguous. If \texttt{cb} is not specified, \texttt{get()} returns an object that can be destructed into the mapping using \texttt{...} notation. For instance, the following looks up in \texttt{VA} using \texttt{T.aid}, and requests the mark's \texttt{x} and \texttt{y} position.  

\begin{example}[Parallel Coordinates]
    \Cref{fig:gallery}(c) visualizes $A.a$ and $B.b$ in $D_1$ on separate axes, and renders $T$ as links.  $V_T$ does not directly map $T.aid$ as in \Cref{fig:gallery}(b). It instead uses $T.aid$ to look up its corresponding point mark via $V_A[aid]$; this follows the foreign key constraint $C_{T\to A}$ and the $1-1$ relationship from $A$ to $V_A$, and then maps the point mark's position (which is a key) to the link's start position.  Referencing $V_A[aid]$ via the foreign keys allows them to be maintained in the visualization even as marks move.   
    {\small\begin{verbatim}  VA = p.dot(A, {y:'a',x:0}), VB = p.dot(B, {y:'b',x:10}) 
  VT = p.link(T, { ...VA.get("aid", {x1:'x',y1:'y'}), 
                   ...VB.get("bid", {x2:'x',y2:'y'})});\end{verbatim}}
\end{example}
\noindent Decomposition $D_2$ only normalizes $b$ so there is a N-1 relationship between $A_2$ and $B_2$.   The following example is often used to render samples of univariate distributions for different conditions:
\begin{example}[Categorical Scatterplot]
    \Cref{fig:gallery}(e) renders $A$ as a scatterplot where $A.bid$ is mapped to x and $V_B$ renders the labels for $B$.  This effectively treats $b$ values as a categorical variable, 
    
    {\small\begin{verbatim}  VA = p.dot(A, { x: 'bid', y: 'a'})
  VB = p.label(B, { ...VA.get('bid','x'), y:-10}) \end{verbatim}}
\end{example}
\noindent Decomposition $D_3$ treats $a$ and $b$ as separate entities with a 1-1 relationship and drops $T$.  The following renders it as a table:
\begin{example}[Table]
     \Cref{fig:gallery}(f) maps $A.a$ (and $B.b$) to labels positioned by their $id$.   $V_A$ and $V_B$ are aligned along the $y$ position so the visualization is faithful.  
     $V_B$'s x-position is the maximum of the right-most edge of the labels in $A$.
     It does this by retrieving $x$ and $w$ of $V_A$ marks, and computing the maximum of \texttt{x+w}.
     
     
    {\small\begin{verbatim} s = p.linear()
 VA = p.label(A, { y:(d)->s(d.id), text:'a', w:'5em', x:0 })
 VB = p.label(B, { y:(d)->s(d.id), text:'b', w:'5em'
   x:VA.get(null,['x','w'],(xws)->max(xws.map(({x,w})->x+w)))}) \end{verbatim}}
\end{example}

\subsubsection{Nesting}

Returning to the hierarchical decomposition $D_2$, we can instead render it as small multiples.  The \texttt{par.nest(child, how?)} API embeds the child view within each mark of the parent. \texttt{how} specifies the path of foreign key relationships, which can be omitted if the path between parent and child is unique.  More generally, \texttt{how} may also be a function that takes as input \texttt{child} and a row in the parent, and returns a table with the same schema as \texttt{child} that will be rendered by the child view.  This simplifies custom nesting logic, such as the second-order logic in \Cref{ss:parallelcords}.
\begin{example}[Small Multiples]
    \Cref{fig:gallery}(d) visualizes $A$ as a scatter plot $V_A$ where the y position is determined by $A.a$, while $B$ is visualized in $V_B$ as rectangles positioned horizontally via  $B.b$.   
    {\small\begin{verbatim}  VB = p.rect(B, { x: 'b', y: 0, w: 10, h: 20})
  VA = p.dot(A, { x: 0, y: 'a'}); VB.nest(VA)   \end{verbatim}}    
    %
\end{example}
\noindent Depending on the definition of individual views, nesting can express faceting, framed-rectangle plots~\cite{cleveland1984graphical}, treemaps, and other nested visualizations.  \Cref{s:hive} describes how a large design space of treemaps are simple to express using \sys.

\section{Compiling HIVE to \sys}\label{s:hive}

HiVE~\cite{hive} is a language to render hierarchical space-filling visualizations.  The developer define a sequence of attributes that should be nested within each other, and specifying at each level the layout algorithm and visual channels to use.  The layout algorithm and how their computed values are mapped to mark properties are baked into the library.   
For instance, the following hierarchically visualizes a Kaggle housing dataset $T(city,type,price,bdrms)$ as a treemap, where the top level is city, followed by housing type.
{\small\begin{verbatim}
  sHier(/,$city,$type);        sLayout(/,SQ,SQ); 
  sSize(/,$price,$bdrms);      sColor(/,_,$bdrms)
\end{verbatim}}
\noindent It uses a squarified layout algorithm that weighs \texttt{city} by \texttt{price} and housing \texttt{type} by \texttt{bdrms}.  The housing types are colored by the number of bedrooms \texttt{bdrms}. 
However, HiVE is limited to one class of visualization designs (hierarchical small-multiples visualizations) and makes strong assumptions about the hierarchical nature of the single input table.  What if these capabilities were available in any visual mapping?
This section  extends our formalism to incorporate space-filling layouts\footnote{Supporting arbitrary layout algorithms is important but beyond our scope here.}, and then describes how HiVE easily compiles to \sys.     

\subsection{Supporting Space-filling Layouts}

HiVE describes seven layout algorithms, including vertical and horizontal partitioning (VT and HZ), and unordered and ordered squarified algorithms~\cite{bruls2000squarified} (SQ and OS).   Each algorithm takes a numeric attribute and a set of rectangular marks as input and computes the size and coordinates for each mark.  For instance, \texttt{SQ(price,V)} ensures each each mark in $V$ is squarish and proportional to $price$ and returns the \texttt{x,y,w,h} values for each mark as an object that maps e.g., \texttt{x} to a numeric vector.

To reference an algorithm, the designer partially applies it with the desired attribute and treats it as a set of data attributes.   For instance, the following renders a table $T$ as rectangle marks by mapping number of bedrooms to color and using \texttt{SQ(type)} to position and size the rectangles; \texttt{...} destructures the return values.
{\small\begin{verbatim}
  p.rect("T", { color: 'bdrms', ...SQ('type')})
\end{verbatim}}
\noindent Internally, \sys partially computes the marks table, passes it to \texttt{SQ()}, and binds the result to x, y, w, and h of the marks.

\subsection{Compilation}
\begin{figure}
    \centering
    \includegraphics[width=0.75\columnwidth]{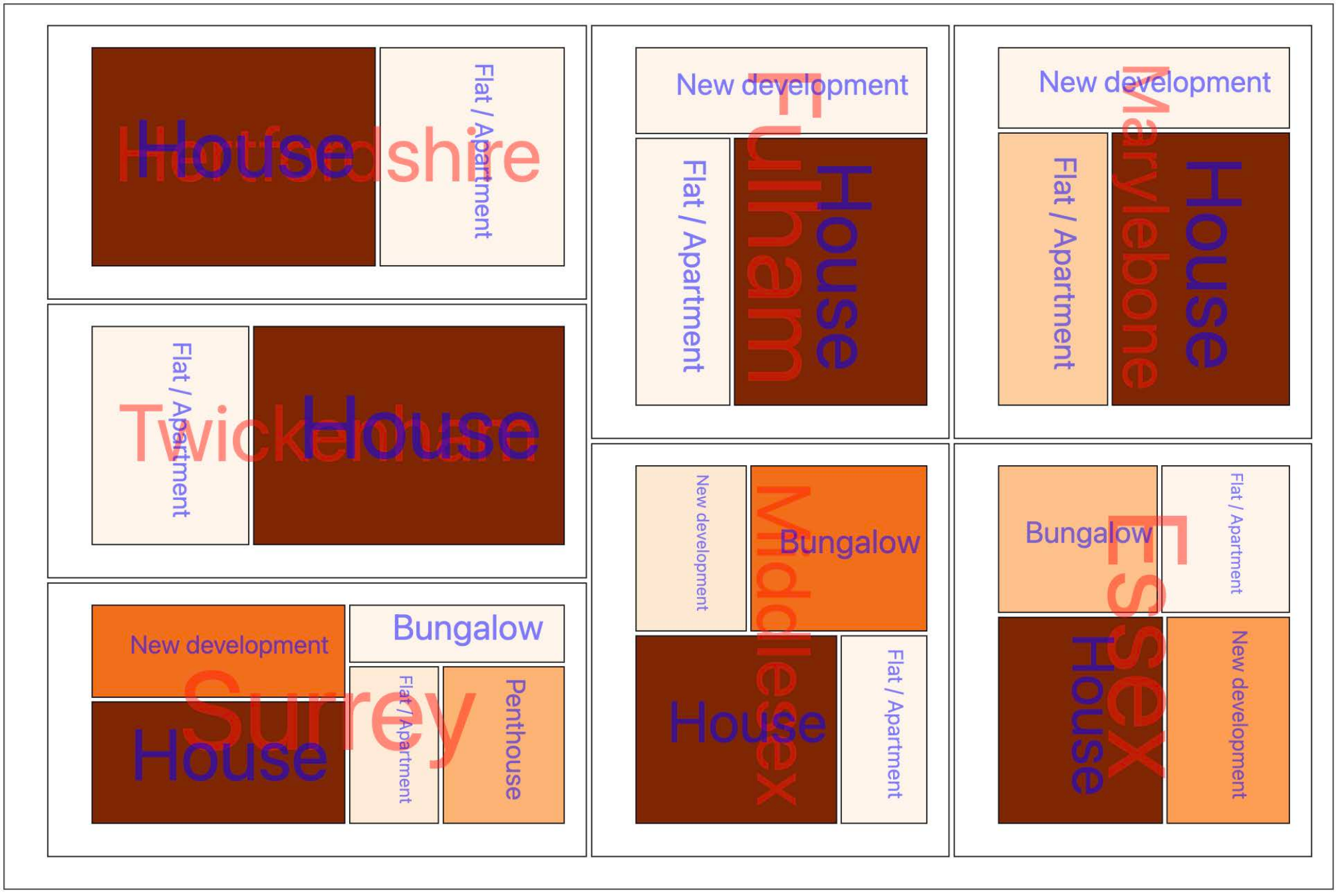}
    \caption{Visualization of nested space-filling visualization.}
    \label{fig:hive}
\end{figure}

%
We use the earlier example to illustrate the process of compiling to \sys.
In the relational model, a hierarchy such as \texttt{sHier} decomposes $T$ into a set of tables that contain each prefix of \texttt{sHier}, as marked in \blue{blue} below.   At each level, the corresponding table also stores the attributes referenced in \texttt{sSize} and \texttt{sColor}.  For instance, $T_c$ stores the prefix \texttt{city} as well as the price because it is referenced in \texttt{sSize}.
{\small\begin{align*}
T_c(\blue{\underline{city}},price),
T_t(\blue{\underline{city,type}},bdrms), \grey{C_1(T_y.type\to\underline{T_t.type})}
\end{align*}}
\noindent \sys provides a hierarchical decomposition function to facilitate this.  It takes as input an attribute hierarchy and optionally a set of additional aggregation expressions to compute at each level of the hierarchy.  The following reproduces the decomposition above:
{\small\begin{verbatim}
  [Tc,Tt] = p.hier("T",  ["city","type"],
      [{price: avg("price")},{bdrms: avg("bdrms")}])
\end{verbatim}}
\noindent Rendering the nested visualization amounts to mapping each table to a set of rectangle marks that maps the squarify or vertical partition algorithms to the rectangle dimensions, and nesting the views appropriately.    Other attributes like color are mapped as normal.  The resulting visualization (label logic omitted) is shown in \Cref{fig:hive}
{\small\begin{verbatim}
   Vc = p.rect(Tc, {...SQ('price')})
   Vt = p.rect(Tt, {...SQ('bdrms'), color:'bdrms'}); 
   Vc.nest(Vt)
\end{verbatim}}
\noindent To summarize, HiVE is readily expressible by hierarchical table decomposition, rectangular mark specifications, and nesting.

\begin{figure}
    \centering
    \includegraphics[width=.75\columnwidth]{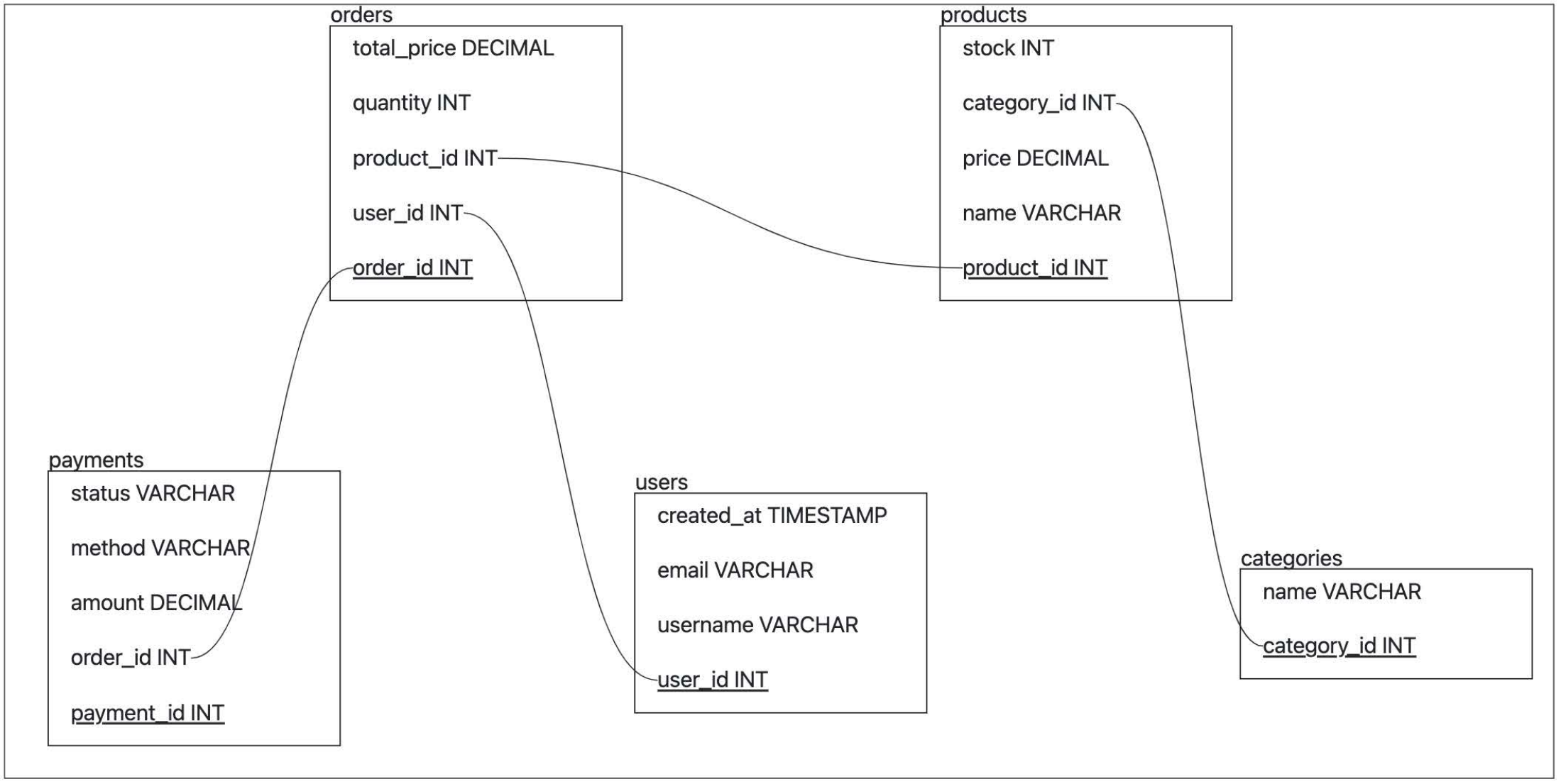}
    \caption{ER Diagram of a shopping platform}
    \label{fig:er}
\end{figure}

\section{Entity Relationship Diagrams}

Entity relationship (ER) diagrams (\Cref{fig:er})  depict the tables and relationships in a database.  They are taught in introductory database courses to aid rapid database design and used in practice to describe a database.   Each table is rendered as a rectangle with the table name at the top, followed by a list of the attributes in the table.  Primary key attributes are underlined, and foreign keys are drawn as edges from the foreign key attributes to the referenced attributes.  ER diagrams are typically rendered in a custom ERD or diagramming tools~\cite{mermaid,dbdiagram,lucid}.  Here, we illustrate how they can be directly expressed over the following database catalog tables. 
{\small \begin{verbatim}
  Tables(id,table_name)
  Columns(id, tid, colname, is_key, type, ord_pos)
  FKeys(id, tid1, col1, tid2, col2)
\end{verbatim}}
\noindent Each column renders as a label, vertically positioned by its ordinal position in the table. Keys are underlined.
{\small \begin{verbatim}
 VC = p.label(Columns, { y: 'ord_pos', x: 0,
    text: ({colname,type})=>`${d.colname} ${d.type}`,
    textDecoration: ({is_key})=>is_key?'underline':'none'})
\end{verbatim}}
\noindent Each table renders as a rectangle whose height is the sum of its column labels and width is the widest column label.  The tables are positioned using a force-directed layout algorithm; edges are passed in as a foreign key reference to \texttt{Fkeys}.   
{\small \begin{verbatim}
  VT = p.rect(Tables, { fill: 'white', stroke: 'black',
    h: Columns.get("id", (rows)=>sum(rows.map(({h})=>h))),
    w: Columns.get("id", (rows)=>max(rows.map(({w})=>w))),
    ...fdlayout(FKeys.get("id", ["tid1","tid2"])) })
\end{verbatim}}
\noindent The table's name is rendered as a label positioned on top of the table, and the columns are nested within the tables.
{\small \begin{verbatim}
  VL = p.label(Tables, { text: 'table_name',
    ...VT.get('id', 'x'),
    y: VT.get('id', 'y', ({y})=>y-10) });   VT.nest(VC)
\end{verbatim}}

\noindent Finally, similar to the node-link diagram in \Cref{fig:nodelink}, we render paths between columns that are related through foreign key references. 
{\small \begin{verbatim}
  VF = p.link(FKeys, { 
    ...VC.get(["tid1","col1"], {x1:"x", y1:"y"}),
    ...VC.get(["tid2","col2"], {x2:"x", y2:"y"})})
\end{verbatim}
}

\begin{figure*}[bt]
    \centering
    \begin{subfigure}[b]{0.28\linewidth}
        \centering
        \includegraphics[width=\linewidth]{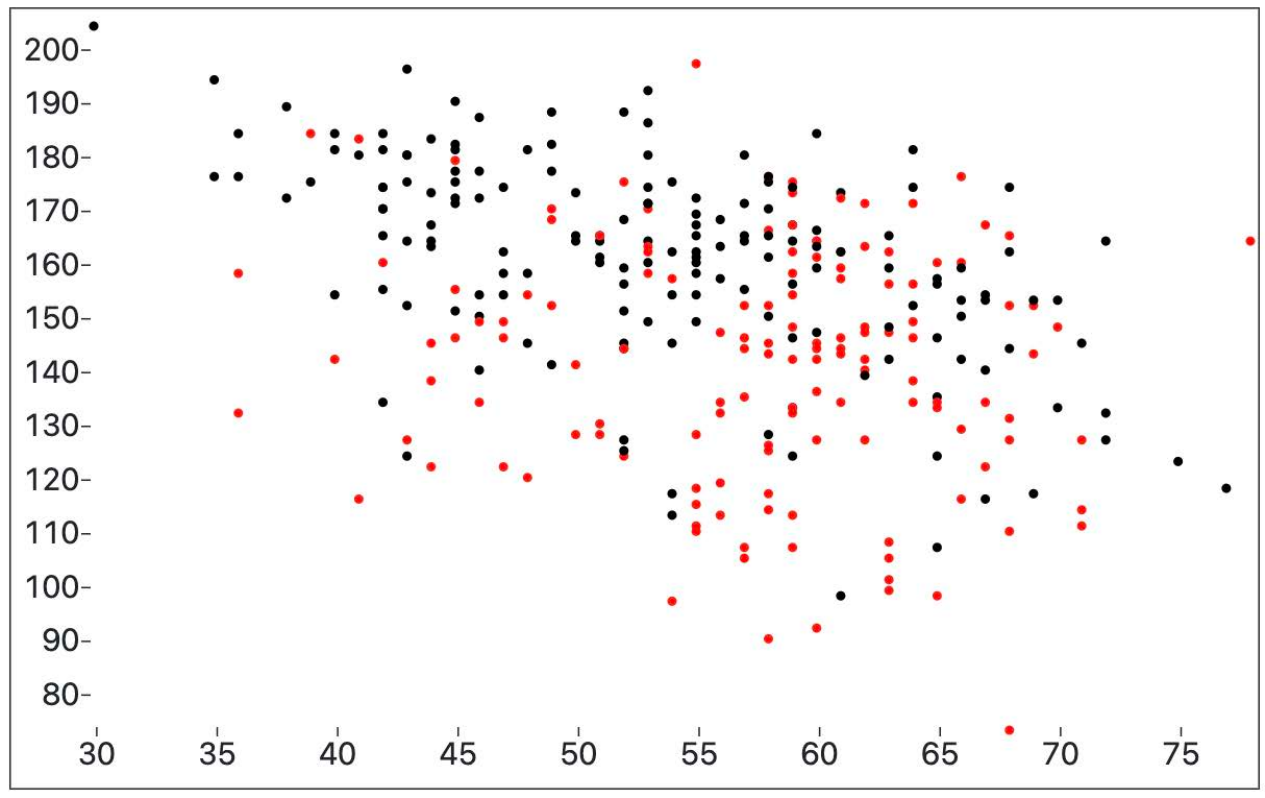}
        \caption{}
        \label{fig:case-scatter(a)}
    \end{subfigure}
    \hfill
    \begin{subfigure}[b]{0.28\linewidth}
        \centering
        \includegraphics[width=\linewidth]{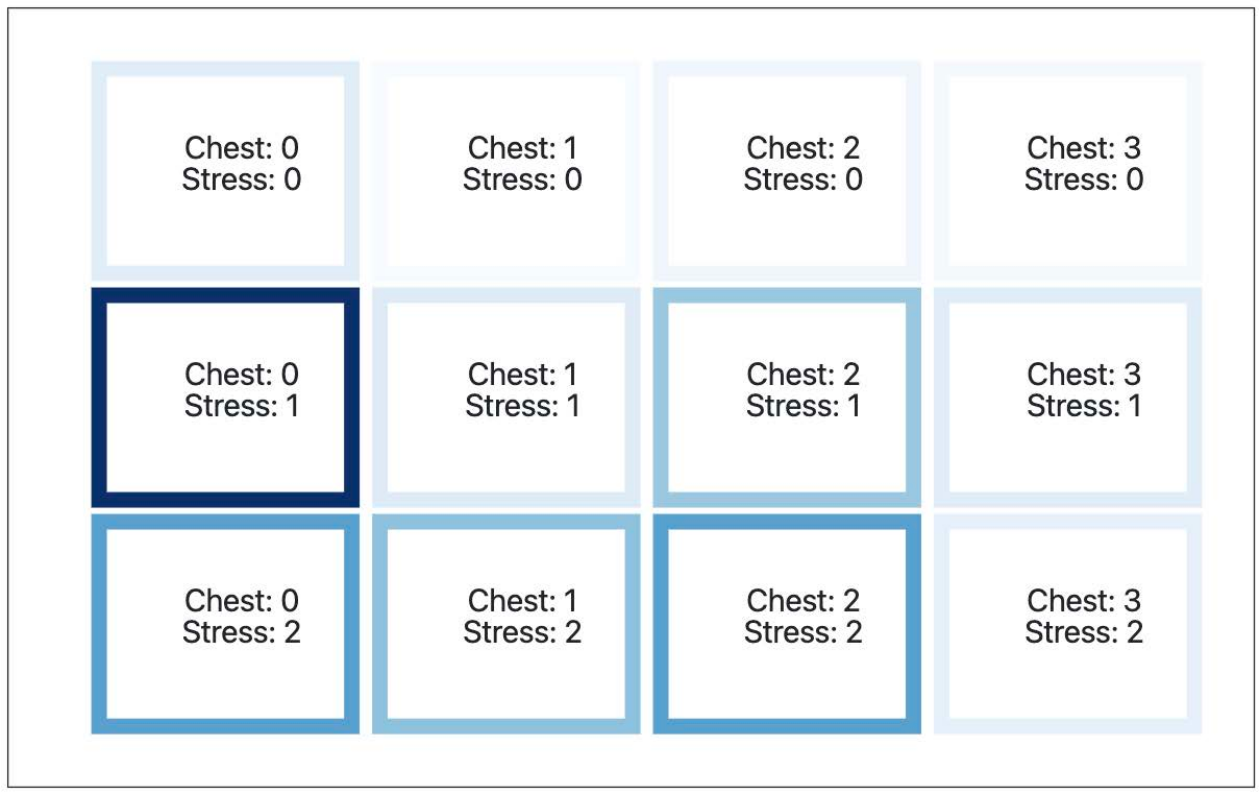}
        \caption{}
        \label{fig:case-scatter(b)}
    \end{subfigure}
    \hfill
   \begin{subfigure}[b]{0.30\linewidth}
       \centering
       \includegraphics[width=\linewidth]{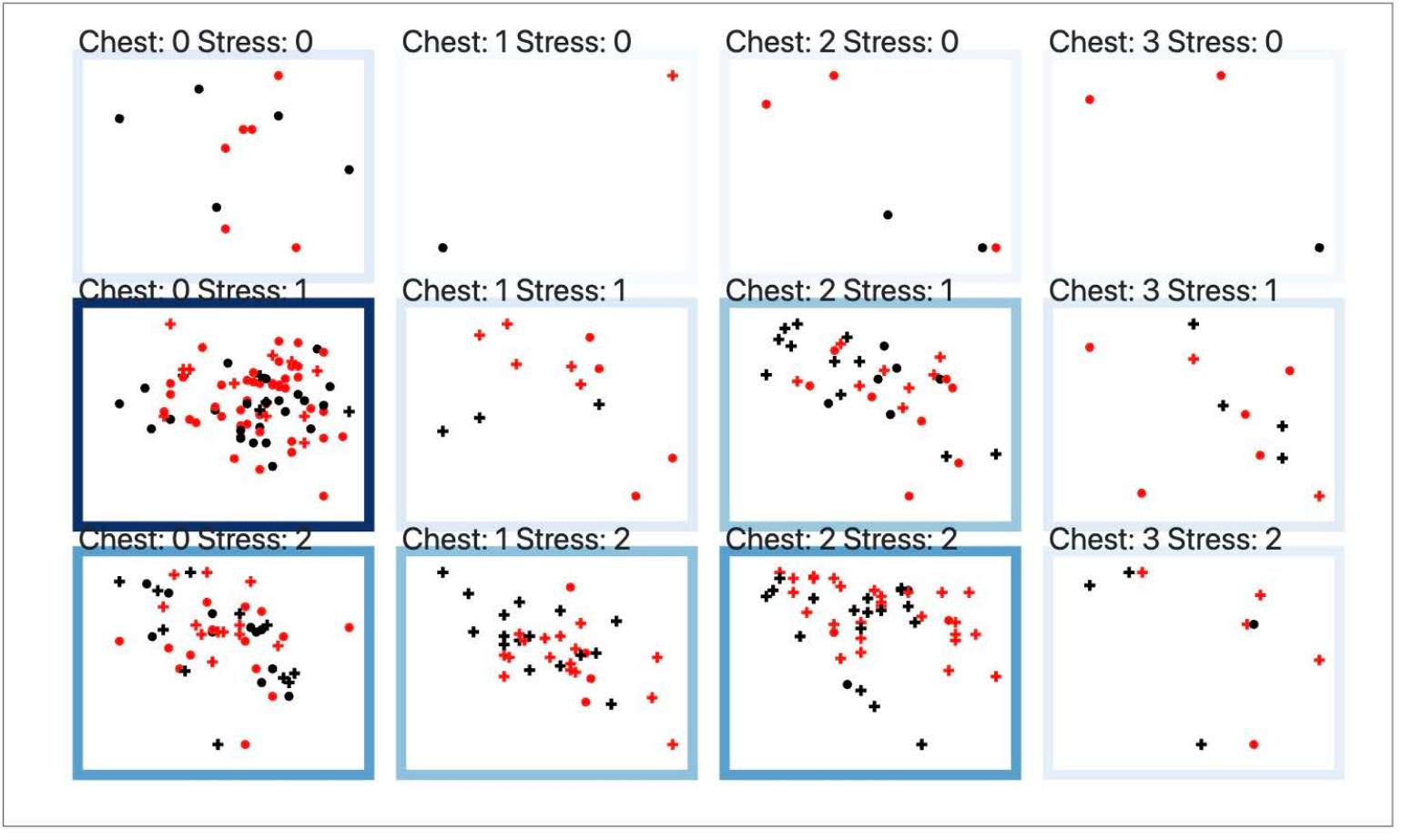}
       \caption{}
       \label{fig:case-scatter(c)}
   \end{subfigure}
    %
    \caption{(a) Scatterplot of age, heart rate and heart disease.  (b) Heatmap visualization of chest pain and heart response to exercise stress.  (c) Scatterplot nested within the heatmap, and the marks with high cholesterol are highlighted in \red{red}.  }
    \label{fig:case-study}
\end{figure*}



%
%

%
%
%

\section{Case Study} 

\label{sec:casestudy}

The following case studies apply \textbf{dvl} to a patient-level cardiovascular 
dataset~\cite{lapp_nd_heartdisease} stored in table \texttt{T}. We show how \textbf{dvl} supports iterative and incremental design with minimal code.
For clarity, we omit code that renders axes.




\subsection{Nested Scatterplots}
We start with a scatterplot in \Cref{fig:case-scatter(a)} that compares age and thalach (the maximum heart rate) when the participant has or doesn't have heart disease (target).  
{\small\begin{verbatim}
    Vd = p.dot(T, { x: 'age', y: 'thalach', symbol: 'target'})
\end{verbatim}}
\noindent \Cref{fig:case-scatter(b)} renders a heat map of chest pain levels (cp) that respond to exercise differently (slope).  The cells are equally sized using the \texttt{EQ()} layout and colored by cardinality.  
{\small\begin{verbatim}
  T2 = T.groupby(['cp', 'slope'], {n: count()})
  Vs = p.rect(T2, {
    x:'cp', y:'slope', stroke: "n", strokeWidth: 10}, 
    {color: {scheme: "blues"}})
\end{verbatim}}


\noindent \sys makes it simple for mark specifications to reference other data and views, which lets designers incrementally build a complex visualization.  For instance, the following renders a label in each cell that is sized to fit inside the cell.  This is done by retrieving the containing cell and calling \texttt{getfontsize(txt, w,h)}, which computes the maximum font size so \texttt{txt} fits within a \texttt{w} by \texttt{h}-pixel box.  Nesting places each label in its corresponding cell.   
{\small\begin{verbatim}
  Vt = p.label(T3, { x:0, y:0, 
    text: ({cp, slope}) => `Chest: ${cp} Stress: ${slope}`,
    fontSize: Vs.get("id", ([{text, w,h}])=> getfontsize(text,w,h))})
  Vs.nest(Vt)
\end{verbatim}}
\noindent This is hard to express in other  libraries.   In Vega-lite~\cite{Satyanarayan2017VegaLiteAG}, the heatmap specification would no longer be applicable.   The designer would need to use create a scatterplot and use faceting, which does not natively support coloring the border to encode the count.

To understand the relationship between \texttt{age} and \texttt{thalach} conditioned on \texttt{cp} and \texttt{slope}, we can easily nest the two views (\Cref{fig:case-scatter(c)}).  \sys infers that the key of \texttt{Vs} is \texttt{cp,slope} and there is a foreign key relationship from \texttt{T.(cp,slope)} to \texttt{Vs.(cp,slope)}. 
We can also highlight participants with cholesterol larger than 230 (e.g., from a slider).   This is expressed by projecting the predicate expression as a new attribute \texttt{sel} mapped to color.   This illustrates that selection interaction is fundamentally a projection operation and does not require specialized mechanisms to express.    We omit code that repositions the labels above their cell.
%
%
{\small\begin{verbatim}
  T4 = T.select({ attrs:"*", sel: gt("chol", 230)})
  Vd = p.dot(T4, {
    x: 'age', y: 'thalach', symbol: 'target', color: 'sel' })
\end{verbatim}}



\subsection{Parallel Coordinates}\label{ss:parallelcords}

\begin{figure}[h]
    \centering
    \includegraphics[width=0.9\linewidth]{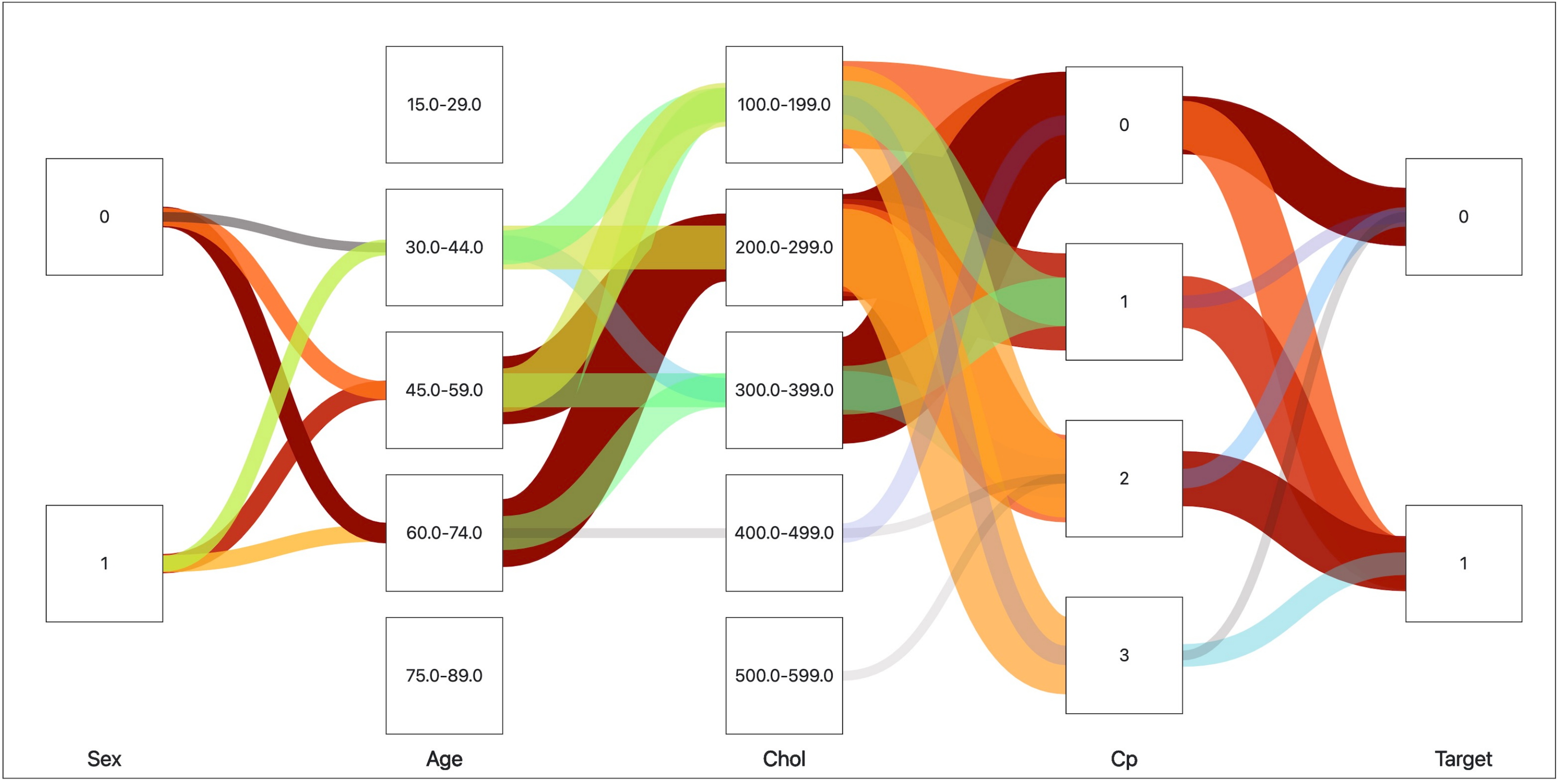} 
    \caption{Parallel coordinate visualization of sex, age, heart rate, chest pain type, and heart disease status, structured through database constraints.}
    \label{fig:parallel_coordinates}
\end{figure}

Parallel coordinates are a way to visualize high-dimensional data by rendering each attribute along parallel axes and drawing each tuple (or group) as paths between the axes. 
We first normalize each attribute in $T$ into a separate attribute table (e.g., \texttt{A}), and define a table \texttt{attrs} that contains the attributes.  We then render equi-spaced rectangles for each attribute that are full height and invisible.  These will serve as containers for each parallel axis.
{\small\begin{verbatim}
  attrs = ["sex", "age", "chol", "cp", "target"]
  tables = T.normalizeMany(attrs) // sex, age, etc tables
  Tattrs = as.table(["attr"], attrs)
  Vatt = p.rect(Tattrs, { ...SQ(), 
    x:"attr", y:0, h:"100%", opacity:"0%"})
\end{verbatim}}
\noindent We then render vertically positioned labels for each attribute table and nest them in \texttt{Vatt}.  The nest function will call \texttt{gv} for each data row in \texttt{Vatt} (e.g., \texttt{{attr:'age'}}), which will return the view for the corresponding attribute (e.g., \texttt{Vage}) that will be nested.
{\small\begin{verbatim}
  views = attrs.map((a,i) =>  p.label(tables[i], {
    y:a, x: 0, text:a, stroke: "black" }))
  gv = (views, {attr})=>views[attrs.indexOf(attr)]
  Vatt.nest(views, gv) 
\end{verbatim}}
\noindent For each row in $T$, we draw curved lines between the marks that render every consecutive pair of attributes.  
{\small\begin{verbatim}
  pairs = zip(attrs.slice(0,-1), attrs.slice(1))
  pairs.map(([a1, a2]) =>
    v1 = gv(views, {attr:a1}), v2 = gv(views, {attr:a2})
    return p.link(T, { 
      x1: v1.get(a1, ({x,w})=>x+w),
      y1: v1.get(a1, ({y,h})=>y+h/2),
      x2: v2.get(a2, "x"),
      y2: v2.get(a2, ({y,h})=>y+h/2) }, {curve: true}))
\end{verbatim}}
\noindent  A comparable visualization in Vega-lite would require $\sim100$ lines of low-level code\footnote{\url{https://vega.github.io/vega-lite/examples/parallel_coordinate.html}}.  Furthermore, a large dataset would typically lead to overplotting, which custom flow diagrams typically address by varying link thickness based on the amount of overplotting.  This is simple to express by changing the above \texttt{return} statement to aggregate by the attributes \texttt{a1,a2} and map the count statistic to link thickness and color (\Cref{fig:parallel_coordinates}).
{\small\begin{verbatim}
  return p.link(T.groupby([a1,a2], {c: count()}), {
      ..., strokeWidth:"c", stroke:"c" })
\end{verbatim}}


\section{Related Work and Discussion}

\subsection{Related Work}

\stitle{Graphical Grammars}
Nearly all visual analysis systems and libraries~\cite{Weaver2004BuildingHV,livny1997devise,north2000snap,derthick1997interactive,aiken1996tioga,Satyanarayan2017VegaLiteAG,wickham2016ggplot2,wilkinson2006grammar,tableau,spotfire,powerbi,sigmacomputing,observable} are built on top of graphical grammars derived from  APT~\cite{mackinlay1986automating},VizQL~\cite{Stolte2000PolarisAS}, grammar of graphics~\cite{wilkinson2012grammar}, and ggplot2~\cite{wickham2016ggplot2}.  As such they all make the single-table assumption and are susceptible to the limitations outlined in \Cref{s:intro}.
Specialized grammars such as HIVE~\cite{hive} are specifically designed to render hierarchical relationships as space filling visualizations.   As we have shown, this is simply a special case of database visualization, with a focus on layout algorithms and nesting.




\stitle{Visualization Constraints.}
Keim's VisDB~\cite{keim2002visdb} uses visualization and data mining techniques to visualize a single table in a way that helps users identify useful patterns such as functional dependencies.  In contrast, \sys is a formalism to map constraints such as functional dependencies to their visual forms.

Algebraic Visualization~\cite{Kindlmann2014AnAP} models the visualization as the output of an algebraic process of data transformation $r$ followed by visual mapping $v$.   This is used to establish invariants when comparing two visualizations.  For instance, a hallucination is when two visual representations ($v_1$, $v_2$) of the same data representation result in different visualizations.  However, algebraic visualization and similar concepts~\cite{bolte2020measures} is an abstract theory and relies on qualitative judgement.  \sys presents an operational definition of faithfulness that can be checked.   Understanding the relationship between these concepts is interesting future work.

Visualization consistency~\cite{qu2017keeping} has been studied in the context of multiple views of the same dataset based on two operational constraints: encode the same data in the same way and encode different data in different ways.   Their focus is on the complex trade-offs practitioners make when it is undesirable for design or perceptual reasons to strictly adhere to all consistency rules.    Similar studies will be important for faithfulness.

%
\subsection{A New Space of Database Visualizations}

Designers and data analysts already spend considerable time preparing, transforming, and munging their data into a single-table form.  While some cleaning and preparation is necessary, some of the work is simply to fit data into a form that visualization tools require.
As weak evidence, data analysis in industry is increasingly consolidating around semantic layers, where data engineers join disparate tables into ``tall-and-wide'' tables and pre-define metrics of interest, and analysts use the table as a single source of truth.  

However, prior works show that pre-joining tables can lead to semantic errors~\cite{Huang2023AggregationCE,Hyde2024MeasuresIS}.  In addition, this work also suggests that visualizations of pre-joined tables are a small subset of the visualization design space.   This implies that the role of the designer changes in two subtle but important ways. 

First, rather than constructing a single table, the designer must decide on a subset of relevant data {\it and} an appropriate decomposition of that data.   While this new-found expressiveness and flexibility potentially places a larger burden on the designer or data engineer, it reduces the concepts that the designer needs to keep track of.   The single-table methodology can still be a default choice, but it is only one in a combinatorial space of valid choices.  

Second, our constraint-oriented formulation means visualizations can violate data constraints.   Thus, the designer's secondary task is to decide how to fix these violations---whether by changing the data, constraints,  decomposition, transformations, or visual mapping---and which violations to allow.

\section{Conclusion and Future Work}

This paper introduced a formalism for database visualization as a mapping from database constraints and contents to a faithful visual representation. 
We started by casting existing visualization formalisms as a mapping from the constraints and contents of a single table to the visualization space, and then extending to foreign-key relationships in order to support relational databases.   
Through examples, we showed the connections between visualization structure and normalization choices.  
Ultimately, database visualization defines a new design space that encapsulates many visualization designs, such as parallel coordinates and facets, that are not expressible under a single-table formalism.

We are excited about several directions of future work.  
The first is to formally incorporate general layout algorithms and to identify the line between declarative mapping specifications and imperative layout algorithms.   To what extent are visualizations ``just queries'' or ``just layout'', really? 
The second is to develop a fast and scalable system to support database visualization.   A particularly interesting aspect is that \sys compiles visualizations into a mixture of SQL and iterative layout algorithms that may be well-suited for GPU database execution.   The third is to extend the theory to general relational constraints (e.g., functional dependencies, check constraints), to non-relational data models (e.g., event streams, tensors), and to incorporate interaction.  The fourth is to understand how this theory can form the basis of a visual data analysis tool that visualizes all data of relevance to the user at once, be they data in a database, a data market, or intermediates in a data science pipeline.


%

%

\bibliographystyle{ACM-Reference-Format}
\bibliography{main}


\begin{thebibliography}{47}


\ifx \showCODEN    \undefined \def \showCODEN     #1{\unskip}     \fi
\ifx \showDOI      \undefined \def \showDOI       #1{#1}\fi
\ifx \showISBNx    \undefined \def \showISBNx     #1{\unskip}     \fi
\ifx \showISBNxiii \undefined \def \showISBNxiii  #1{\unskip}     \fi
\ifx \showISSN     \undefined \def \showISSN      #1{\unskip}     \fi
\ifx \showLCCN     \undefined \def \showLCCN      #1{\unskip}     \fi
\ifx \shownote     \undefined \def \shownote      #1{#1}          \fi
\ifx \showarticletitle \undefined \def \showarticletitle #1{#1}   \fi
\ifx \showURL      \undefined \def \showURL       {\relax}        \fi
\providecommand\bibfield[2]{#2}
\providecommand\bibinfo[2]{#2}
\providecommand\natexlab[1]{#1}
\providecommand\showeprint[2][]{arXiv:#2}

\bibitem[dbd({[n.\,d.]})]%
        {dbdiagram}
 \bibinfo{year}{[n.\,d.]}\natexlab{}.
\newblock \bibinfo{title}{Draw Entity-Relationship Diagrams, Painlessly}.
\newblock \bibinfo{howpublished}{\url{https://dbdiagram.io/home}}.
\newblock


\bibitem[luc({[n.\,d.]})]%
        {lucid}
 \bibinfo{year}{[n.\,d.]}\natexlab{}.
\newblock \bibinfo{title}{Lucid Charts}.
\newblock \bibinfo{howpublished}{\url{https://www.lucidchart.com/pages/examples/er-diagram-tool}}.
\newblock


\bibitem[mer({[n.\,d.]})]%
        {mermaid}
 \bibinfo{year}{[n.\,d.]}\natexlab{}.
\newblock \bibinfo{title}{Mermaid Diagramming and charting tool}.
\newblock \bibinfo{howpublished}{\url{https://mermaid.js.org/}}.
\newblock


\bibitem[obs(2024)]%
        {observable}
 \bibinfo{year}{2024}\natexlab{}.
\newblock \bibinfo{title}{Observable: The best data visualizations are built with code.}
\newblock \bibinfo{howpublished}{\url{https://observablehq.com/}}.
\newblock


\bibitem[Aiken et~al\mbox{.}(1996)]%
        {aiken1996tioga}
\bibfield{author}{\bibinfo{person}{Alexander Aiken}, \bibinfo{person}{Jolly Chen}, \bibinfo{person}{Michael Stonebraker}, {and} \bibinfo{person}{Allison Woodruff}.} \bibinfo{year}{1996}\natexlab{}.
\newblock \showarticletitle{Tioga-2: A Direct Manipulation Database Visualization Environment}. In \bibinfo{booktitle}{\emph{ICDE}}. \bibinfo{pages}{208--217}.
\newblock


\bibitem[Bertin(1983)]%
        {Bertin1983TheSO}
\bibfield{author}{\bibinfo{person}{John~J. Bertin}.} \bibinfo{year}{1983}\natexlab{}.
\newblock \showarticletitle{The semiology of graphics}.
\newblock
\urldef\tempurl%
\url{https://api.semanticscholar.org/CorpusID:59708204}
\showURL{%
\tempurl}


\bibitem[Bolte and Bruckner(2020)]%
        {bolte2020measures}
\bibfield{author}{\bibinfo{person}{Fabian Bolte} {and} \bibinfo{person}{Stefan Bruckner}.} \bibinfo{year}{2020}\natexlab{}.
\newblock \showarticletitle{Measures in visualization space}.
\newblock In \bibinfo{booktitle}{\emph{Foundations of Data Visualization}}. \bibinfo{publisher}{Springer}, \bibinfo{pages}{39--59}.
\newblock


\bibitem[Bruls et~al\mbox{.}(2000)]%
        {bruls2000squarified}
\bibfield{author}{\bibinfo{person}{Mark Bruls}, \bibinfo{person}{Kees Huizing}, {and} \bibinfo{person}{Jarke~J Van~Wijk}.} \bibinfo{year}{2000}\natexlab{}.
\newblock \showarticletitle{Squarified treemaps}. In \bibinfo{booktitle}{\emph{Data Visualization 2000: Proceedings of the Joint EUROGRAPHICS and IEEE TCVG Symposium on Visualization in Amsterdam, The Netherlands, May 29--30, 2000}}. Springer, \bibinfo{pages}{33--42}.
\newblock


\bibitem[Card et~al\mbox{.}(1999)]%
        {card1999readings}
\bibfield{author}{\bibinfo{person}{Stuart~K Card}, \bibinfo{person}{Jock Mackinlay}, {and} \bibinfo{person}{Ben Shneiderman}.} \bibinfo{year}{1999}\natexlab{}.
\newblock \bibinfo{booktitle}{\emph{Readings in information visualization: using vision to think}}.
\newblock \bibinfo{publisher}{Morgan Kaufmann}.
\newblock


\bibitem[Chen and Liu(2023)]%
        {chen2023state}
\bibfield{author}{\bibinfo{person}{Chen Chen} {and} \bibinfo{person}{Zhicheng Liu}.} \bibinfo{year}{2023}\natexlab{}.
\newblock \showarticletitle{The state of the art in creating visualization corpora for automated chart analysis}. In \bibinfo{booktitle}{\emph{Computer Graphics Forum}}, Vol.~\bibinfo{volume}{42}. Wiley Online Library, \bibinfo{pages}{449--470}.
\newblock


\bibitem[Cleveland and McGill(1984)]%
        {cleveland1984graphical}
\bibfield{author}{\bibinfo{person}{William~S Cleveland} {and} \bibinfo{person}{Robert McGill}.} \bibinfo{year}{1984}\natexlab{}.
\newblock \showarticletitle{Graphical Perception: Theory, Experimentation, and Application to the Development Of Graphical Methods}.
\newblock \bibinfo{journal}{\emph{Journal of the American statistical association}}  \bibinfo{volume}{79} (\bibinfo{year}{1984}), \bibinfo{pages}{531--554}.
\newblock


\bibitem[Codd(1970)]%
        {codd1970relational}
\bibfield{author}{\bibinfo{person}{Edgar~F Codd}.} \bibinfo{year}{1970}\natexlab{}.
\newblock \showarticletitle{A relational model of data for large shared data banks}.
\newblock \bibinfo{journal}{\emph{Commun. ACM}} \bibinfo{volume}{13}, \bibinfo{number}{6} (\bibinfo{year}{1970}), \bibinfo{pages}{377--387}.
\newblock


\bibitem[Derthick et~al\mbox{.}(1997)]%
        {derthick1997interactive}
\bibfield{author}{\bibinfo{person}{Mark Derthick}, \bibinfo{person}{John Kolojejchick}, {and} \bibinfo{person}{Steven~F Roth}.} \bibinfo{year}{1997}\natexlab{}.
\newblock \showarticletitle{An Interactive Visualization Environment for Data Exploration.}. In \bibinfo{booktitle}{\emph{KDD}}. \bibinfo{pages}{2--9}.
\newblock


\bibitem[Gibson et~al\mbox{.}(2013)]%
        {gibson2013survey}
\bibfield{author}{\bibinfo{person}{Helen Gibson}, \bibinfo{person}{Joe Faith}, {and} \bibinfo{person}{Paul Vickers}.} \bibinfo{year}{2013}\natexlab{}.
\newblock \showarticletitle{A survey of two-dimensional graph layout techniques for information visualisation}.
\newblock \bibinfo{journal}{\emph{Information visualization}} \bibinfo{volume}{12}, \bibinfo{number}{3-4} (\bibinfo{year}{2013}), \bibinfo{pages}{324--357}.
\newblock


\bibitem[Huang et~al\mbox{.}(2023)]%
        {Huang2023AggregationCE}
\bibfield{author}{\bibinfo{person}{Zezhou Huang}, \bibinfo{person}{Pavan~Kalyan Damalapati}, {and} \bibinfo{person}{Eugene Wu}.} \bibinfo{year}{2023}\natexlab{}.
\newblock \showarticletitle{Aggregation Consistency Errors in Semantic Layers and How to Avoid Them}.
\newblock \bibinfo{journal}{\emph{Proceedings of the Workshop on Human-In-the-Loop Data Analytics}} (\bibinfo{year}{2023}).
\newblock
\urldef\tempurl%
\url{https://api.semanticscholar.org/CorpusID:259316670}
\showURL{%
\tempurl}


\bibitem[Hyde and Fremlin(2024)]%
        {Hyde2024MeasuresIS}
\bibfield{author}{\bibinfo{person}{Julian Hyde} {and} \bibinfo{person}{John Fremlin}.} \bibinfo{year}{2024}\natexlab{}.
\newblock \showarticletitle{Measures in SQL}.
\newblock \bibinfo{journal}{\emph{Companion of the 2024 International Conference on Management of Data}} (\bibinfo{year}{2024}).
\newblock
\urldef\tempurl%
\url{https://api.semanticscholar.org/CorpusID:269987193}
\showURL{%
\tempurl}


\bibitem[Keim and Kriegel(2002)]%
        {keim2002visdb}
\bibfield{author}{\bibinfo{person}{Daniel~A Keim} {and} \bibinfo{person}{H-P Kriegel}.} \bibinfo{year}{2002}\natexlab{}.
\newblock \showarticletitle{VisDB: Database exploration using multidimensional visualization}.
\newblock \bibinfo{journal}{\emph{IEEE Computer Graphics and Applications}} \bibinfo{volume}{14}, \bibinfo{number}{5} (\bibinfo{year}{2002}), \bibinfo{pages}{40--49}.
\newblock


\bibitem[Kindlmann and Scheidegger(2014)]%
        {Kindlmann2014AnAP}
\bibfield{author}{\bibinfo{person}{Gordon~L. Kindlmann} {and} \bibinfo{person}{Carlos~Eduardo Scheidegger}.} \bibinfo{year}{2014}\natexlab{}.
\newblock \showarticletitle{An Algebraic Process for Visualization Design}.
\newblock \bibinfo{journal}{\emph{IEEE Transactions on Visualization and Computer Graphics}}  \bibinfo{volume}{20} (\bibinfo{year}{2014}), \bibinfo{pages}{2181--2190}.
\newblock
\urldef\tempurl%
\url{https://api.semanticscholar.org/CorpusID:10535011}
\showURL{%
\tempurl}


\bibitem[Kohn et~al\mbox{.}(2022)]%
        {kohn2022duckdb}
\bibfield{author}{\bibinfo{person}{Andr{\'e} Kohn}, \bibinfo{person}{Dominik Moritz}, \bibinfo{person}{Mark Raasveldt}, \bibinfo{person}{Hannes M{\"u}hleisen}, {and} \bibinfo{person}{Thomas Neumann}.} \bibinfo{year}{2022}\natexlab{}.
\newblock \showarticletitle{DuckDB-wasm: fast analytical processing for the web}.
\newblock \bibinfo{journal}{\emph{Proceedings of the VLDB Endowment}} \bibinfo{volume}{15}, \bibinfo{number}{12} (\bibinfo{year}{2022}), \bibinfo{pages}{3574--3577}.
\newblock


\bibitem[Lapp(nd)]%
        {lapp_nd_heartdisease}
\bibfield{author}{\bibinfo{person}{David Lapp}.} \bibinfo{year}{n.d.}\natexlab{}.
\newblock \bibinfo{title}{Heart Disease Dataset}.
\newblock \bibinfo{howpublished}{Kaggle}.
\newblock
\urldef\tempurl%
\url{https://www.kaggle.com/datasets/johnsmith88/heart-disease-dataset}
\showURL{%
\tempurl}


\bibitem[Lausen et~al\mbox{.}(1998)]%
        {lausen1998logical}
\bibfield{author}{\bibinfo{person}{Georg Lausen}, \bibinfo{person}{Bertram Lud{\"a}scher}, {and} \bibinfo{person}{Wolfgang May}.} \bibinfo{year}{1998}\natexlab{}.
\newblock \showarticletitle{On logical foundations of active databases}.
\newblock In \bibinfo{booktitle}{\emph{Logics for Databases and Information Systems}}. \bibinfo{publisher}{Springer}, \bibinfo{pages}{389--422}.
\newblock


\bibitem[Livny et~al\mbox{.}(1997)]%
        {livny1997devise}
\bibfield{author}{\bibinfo{person}{Miron Livny}, \bibinfo{person}{Raghu Ramakrishnan}, \bibinfo{person}{Kevin Beyer}, \bibinfo{person}{Guangshun Chen}, \bibinfo{person}{Donko Donjerkovic}, \bibinfo{person}{Shilpa Lawande}, \bibinfo{person}{Jussi Myllymaki}, {and} \bibinfo{person}{Kent Wenger}.} \bibinfo{year}{1997}\natexlab{}.
\newblock \showarticletitle{Devise: Integrated Querying and Visual Exploration Of Large Datasets}.
\newblock \bibinfo{journal}{\emph{ACM SIGMOD Record}} \bibinfo{volume}{26}, \bibinfo{number}{2} (\bibinfo{year}{1997}), \bibinfo{pages}{301--312}.
\newblock


\bibitem[Mackinlay(1986)]%
        {mackinlay1986automating}
\bibfield{author}{\bibinfo{person}{Jock Mackinlay}.} \bibinfo{year}{1986}\natexlab{}.
\newblock \showarticletitle{Automating the design of graphical presentations of relational information}.
\newblock \bibinfo{journal}{\emph{Acm Transactions On Graphics (Tog)}} \bibinfo{volume}{5}, \bibinfo{number}{2} (\bibinfo{year}{1986}), \bibinfo{pages}{110--141}.
\newblock


\bibitem[Mackinlay et~al\mbox{.}(2007)]%
        {mackinlay2007show}
\bibfield{author}{\bibinfo{person}{Jock Mackinlay}, \bibinfo{person}{Pat Hanrahan}, {and} \bibinfo{person}{Chris Stolte}.} \bibinfo{year}{2007}\natexlab{}.
\newblock \showarticletitle{Show me: Automatic presentation for visual analysis}.
\newblock \bibinfo{journal}{\emph{IEEE transactions on visualization and computer graphics}} \bibinfo{volume}{13}, \bibinfo{number}{6} (\bibinfo{year}{2007}), \bibinfo{pages}{1137--1144}.
\newblock


\bibitem[Mcnutt(2022)]%
        {Mcnutt2022NoGT}
\bibfield{author}{\bibinfo{person}{Andrew~M Mcnutt}.} \bibinfo{year}{2022}\natexlab{}.
\newblock \showarticletitle{No Grammar to Rule Them All: A Survey of JSON-style DSLs for Visualization}.
\newblock \bibinfo{journal}{\emph{IEEE Transactions on Visualization and Computer Graphics}}  \bibinfo{volume}{29} (\bibinfo{year}{2022}), \bibinfo{pages}{160--170}.
\newblock
\urldef\tempurl%
\url{https://api.semanticscholar.org/CorpusID:250627665}
\showURL{%
\tempurl}


\bibitem[Mcnutt and Kindlmann(2018)]%
        {Mcnutt2018LintingFV}
\bibfield{author}{\bibinfo{person}{Andrew~M Mcnutt} {and} \bibinfo{person}{Gordon~L. Kindlmann}.} \bibinfo{year}{2018}\natexlab{}.
\newblock \showarticletitle{Linting for Visualization: Towards a Practical Automated Visualization Guidance System}.
\newblock
\urldef\tempurl%
\url{https://api.semanticscholar.org/CorpusID:208615818}
\showURL{%
\tempurl}


\bibitem[Moritz et~al\mbox{.}(2019)]%
        {2019-draco}
\bibfield{author}{\bibinfo{person}{Dominik Moritz}, \bibinfo{person}{Chenglong Wang}, \bibinfo{person}{Gregory Nelson}, \bibinfo{person}{Halden Lin}, \bibinfo{person}{Adam~M. Smith}, \bibinfo{person}{Bill Howe}, {and} \bibinfo{person}{Jeffrey Heer}.} \bibinfo{year}{2019}\natexlab{}.
\newblock \showarticletitle{Formalizing Visualization Design Knowledge as Constraints: Actionable and Extensible Models in Draco}.
\newblock \bibinfo{journal}{\emph{IEEE Trans. Visualization \& Comp. Graphics (Proc. InfoVis)}} (\bibinfo{year}{2019}).
\newblock
\urldef\tempurl%
\url{http://idl.cs.washington.edu/papers/draco}
\showURL{%
\tempurl}


\bibitem[North and Shneiderman(2000)]%
        {north2000snap}
\bibfield{author}{\bibinfo{person}{Chris North} {and} \bibinfo{person}{Ben Shneiderman}.} \bibinfo{year}{2000}\natexlab{}.
\newblock \showarticletitle{Snap-Together Visualization: a User Interface for Coordinating Visualizations Via Relational Schemata}. In \bibinfo{booktitle}{\emph{AVI}}. \bibinfo{pages}{128--135}.
\newblock


\bibitem[powerbi({[n.\,d.]})]%
        {powerbi}
powerbi \bibinfo{year}{[n.\,d.]}\natexlab{}.
\newblock \bibinfo{title}{Microsoft Powerbi}.
\newblock \bibinfo{howpublished}{\url{https://powerbi.microsoft.com/en-us/}}.
\newblock


\bibitem[Qu and Hullman(2017)]%
        {qu2017keeping}
\bibfield{author}{\bibinfo{person}{Zening Qu} {and} \bibinfo{person}{Jessica Hullman}.} \bibinfo{year}{2017}\natexlab{}.
\newblock \showarticletitle{Keeping multiple views consistent: Constraints, validations, and exceptions in visualization authoring}.
\newblock \bibinfo{journal}{\emph{IEEE transactions on visualization and computer graphics}} \bibinfo{volume}{24}, \bibinfo{number}{1} (\bibinfo{year}{2017}), \bibinfo{pages}{468--477}.
\newblock


\bibitem[Satyanarayan et~al\mbox{.}(2016)]%
        {Satyanarayan2017VegaLiteAG}
\bibfield{author}{\bibinfo{person}{Arvind Satyanarayan}, \bibinfo{person}{Dominik Moritz}, \bibinfo{person}{Kanit Wongsuphasawat}, {and} \bibinfo{person}{J. Heer}.} \bibinfo{year}{2016}\natexlab{}.
\newblock \showarticletitle{Vega-Lite: A Grammar Of Interactive Graphics}.
\newblock \bibinfo{journal}{\emph{TVCG}} \bibinfo{volume}{23}, \bibinfo{number}{1} (\bibinfo{year}{2016}), \bibinfo{pages}{341--350}.
\newblock


\bibitem[Satyanarayan et~al\mbox{.}(2018)]%
        {Satyanarayan2018VegaLiteAG}
\bibfield{author}{\bibinfo{person}{Arvind Satyanarayan}, \bibinfo{person}{Dominik Moritz}, \bibinfo{person}{Kanit Wongsuphasawat}, {and} \bibinfo{person}{Jeffrey Heer}.} \bibinfo{year}{2018}\natexlab{}.
\newblock \showarticletitle{Vega-Lite: A Grammar of Interactive Graphics}.
\newblock \bibinfo{journal}{\emph{IEEE Transactions on Visualization and Computer Graphics}}  \bibinfo{volume}{23} (\bibinfo{year}{2018}), \bibinfo{pages}{341--350}.
\newblock
\urldef\tempurl%
\url{https://api.semanticscholar.org/CorpusID:206805969}
\showURL{%
\tempurl}


\bibitem[Schulz et~al\mbox{.}(2010)]%
        {schulz2010design}
\bibfield{author}{\bibinfo{person}{Hans-Jorg Schulz}, \bibinfo{person}{Steffen Hadlak}, {and} \bibinfo{person}{Heidrun Schumann}.} \bibinfo{year}{2010}\natexlab{}.
\newblock \showarticletitle{The design space of implicit hierarchy visualization: A survey}.
\newblock \bibinfo{journal}{\emph{IEEE transactions on visualization and computer graphics}} \bibinfo{volume}{17}, \bibinfo{number}{4} (\bibinfo{year}{2010}), \bibinfo{pages}{393--411}.
\newblock


\bibitem[sigmacomputing({[n.\,d.]})]%
        {sigmacomputing}
sigmacomputing \bibinfo{year}{[n.\,d.]}\natexlab{}.
\newblock \bibinfo{title}{Sigma Computing}.
\newblock \bibinfo{howpublished}{\url{https://www.sigmacomputing.com/}}.
\newblock


\bibitem[Slingsby et~al\mbox{.}(2009)]%
        {hive}
\bibfield{author}{\bibinfo{person}{Aidan Slingsby}, \bibinfo{person}{Jason Dykes}, {and} \bibinfo{person}{Jo Wood}.} \bibinfo{year}{2009}\natexlab{}.
\newblock \showarticletitle{Configuring hierarchical layouts to address research questions}.
\newblock \bibinfo{journal}{\emph{IEEE transactions on visualization and computer graphics}} \bibinfo{volume}{15}, \bibinfo{number}{6} (\bibinfo{year}{2009}), \bibinfo{pages}{977--984}.
\newblock


\bibitem[spotfire({[n.\,d.]})]%
        {spotfire}
spotfire \bibinfo{year}{[n.\,d.]}\natexlab{}.
\newblock \bibinfo{title}{Tibco Spotfire}.
\newblock \bibinfo{howpublished}{\url{https://www.tibco.com/products/tibco-spotfire}}.
\newblock


\bibitem[Stolte and Hanrahan(2002)]%
        {Stolte2000PolarisAS}
\bibfield{author}{\bibinfo{person}{C. Stolte} {and} \bibinfo{person}{P. Hanrahan}.} \bibinfo{year}{2002}\natexlab{}.
\newblock \showarticletitle{Polaris: a System for Query, Analysis and Visualization Of Multi-Dimensional Relational Databases}.
\newblock \bibinfo{journal}{\emph{TVCG}}  \bibinfo{volume}{8} (\bibinfo{year}{2002}), \bibinfo{pages}{52--65}.
\newblock


\bibitem[Stolte et~al\mbox{.}(2002)]%
        {stolte2002polaris}
\bibfield{author}{\bibinfo{person}{Chris Stolte}, \bibinfo{person}{Diane Tang}, {and} \bibinfo{person}{Pat Hanrahan}.} \bibinfo{year}{2002}\natexlab{}.
\newblock \showarticletitle{Polaris: A system for query, analysis, and visualization of multidimensional relational databases}.
\newblock \bibinfo{journal}{\emph{IEEE Transactions on Visualization and Computer Graphics}} \bibinfo{volume}{8}, \bibinfo{number}{1} (\bibinfo{year}{2002}), \bibinfo{pages}{52--65}.
\newblock


\bibitem[tableau({[n.\,d.]})]%
        {tableau}
tableau \bibinfo{year}{[n.\,d.]}\natexlab{}.
\newblock \bibinfo{title}{Tableau Software}.
\newblock \bibinfo{howpublished}{\url{http://www.tableau.com}}.
\newblock


\bibitem[Talbot et~al\mbox{.}(2014)]%
        {talbot2014four}
\bibfield{author}{\bibinfo{person}{Justin Talbot}, \bibinfo{person}{Vidya Setlur}, {and} \bibinfo{person}{Anushka Anand}.} \bibinfo{year}{2014}\natexlab{}.
\newblock \showarticletitle{Four experiments on the perception of bar charts}.
\newblock \bibinfo{journal}{\emph{IEEE transactions on visualization and computer graphics}} \bibinfo{volume}{20}, \bibinfo{number}{12} (\bibinfo{year}{2014}), \bibinfo{pages}{2152--2160}.
\newblock


\bibitem[Weaver(2004)]%
        {Weaver2004BuildingHV}
\bibfield{author}{\bibinfo{person}{Chris Weaver}.} \bibinfo{year}{2004}\natexlab{}.
\newblock \showarticletitle{Building Highly-Coordinated Visualizations in Improvise}.
\newblock \bibinfo{journal}{\emph{IEEE Symposium on Information Visualization}} (\bibinfo{year}{2004}), \bibinfo{pages}{159--166}.
\newblock


\bibitem[Wickham(2010)]%
        {Wickham2010ALG}
\bibfield{author}{\bibinfo{person}{Hadley Wickham}.} \bibinfo{year}{2010}\natexlab{}.
\newblock \showarticletitle{A Layered Grammar of Graphics}.
\newblock \bibinfo{journal}{\emph{Journal of Computational and Graphical Statistics}}  \bibinfo{volume}{19} (\bibinfo{year}{2010}), \bibinfo{pages}{28 -- 3}.
\newblock
\urldef\tempurl%
\url{https://api.semanticscholar.org/CorpusID:266499321}
\showURL{%
\tempurl}


\bibitem[Wickham(2016)]%
        {wickham2016ggplot2}
\bibfield{author}{\bibinfo{person}{Hadley Wickham}.} \bibinfo{year}{2016}\natexlab{}.
\newblock \bibinfo{booktitle}{\emph{Ggplot2: Elegant Graphics for Data Analysis}}.
\newblock \bibinfo{publisher}{Springer}.
\newblock


\bibitem[Wilkinson(2001)]%
        {Wilkinson2001nViZnA}
\bibfield{author}{\bibinfo{person}{Leland Wilkinson}.} \bibinfo{year}{2001}\natexlab{}.
\newblock \showarticletitle{nViZn : An Algebra-Based Visualization System}.
\newblock
\urldef\tempurl%
\url{https://api.semanticscholar.org/CorpusID:14945586}
\showURL{%
\tempurl}


\bibitem[Wilkinson(2006)]%
        {wilkinson2006grammar}
\bibfield{author}{\bibinfo{person}{Leland Wilkinson}.} \bibinfo{year}{2006}\natexlab{}.
\newblock \bibinfo{booktitle}{\emph{The Grammar Of Graphics}}.
\newblock \bibinfo{publisher}{Springer Science \& Business Media}.
\newblock


\bibitem[Wilkinson(2012)]%
        {wilkinson2012grammar}
\bibfield{author}{\bibinfo{person}{Leland Wilkinson}.} \bibinfo{year}{2012}\natexlab{}.
\newblock \bibinfo{booktitle}{\emph{The grammar of graphics}}.
\newblock \bibinfo{publisher}{Springer}.
\newblock


\bibitem[Wongsuphasawat et~al\mbox{.}(2015)]%
        {wongsuphasawat2015voyager}
\bibfield{author}{\bibinfo{person}{Kanit Wongsuphasawat}, \bibinfo{person}{Dominik Moritz}, \bibinfo{person}{Anushka Anand}, \bibinfo{person}{Jock Mackinlay}, \bibinfo{person}{Bill Howe}, {and} \bibinfo{person}{Jeffrey Heer}.} \bibinfo{year}{2015}\natexlab{}.
\newblock \showarticletitle{Voyager: Exploratory analysis via faceted browsing of visualization recommendations}.
\newblock \bibinfo{journal}{\emph{IEEE transactions on visualization and computer graphics}} \bibinfo{volume}{22}, \bibinfo{number}{1} (\bibinfo{year}{2015}), \bibinfo{pages}{649--658}.
\newblock


\end{thebibliography}

\end{document}